\documentclass[useAMS,usenatbib]{mn2e}
\usepackage{epsfig}

\title[Optical Variability of TeV Blazars]{Optical Variability of TeV Blazars on long time-scales}

\author[Gaur et al.]
{Haritma Gaur$^{1}$\thanks{E-mail: haritma@aries.res.in}, Alok C.\ Gupta$^{1}$, R.\ Bachev$^{2}$, A.\ Strigachev$^{2}$, 
E.\ Semkov$^{2}$, 
\newauthor P. J.\ Wiita$^{3}$, O. M. Kurtanidze$^{4,5,6,7}$, A. Darriba$^{8,9}$, G. Damljanovic$^{10}$, 
\newauthor R. G. Chanishvili$^{4}$, S. Ibryamov$^{11}$, S. O. Kurtanidze$^{4}$, M. G. Nikolashvili$^{4}$, 
\newauthor L. A. Sigua$^{4}$, O. Vince$^{10}$   \\
\\
$^{1}$Aryabhatta Research Institute of Observational Sciences (ARIES), Manora Peak, Nainital -- 263002, India \\
$^{2}$Institute of Astronomy and National Astronomical Observatory,Bulgarian Academy of Sciences, \\
~~72 Tsarigradsko Shosse Blvd., 1784 Sofia, Bulgaria  \\
$^{3}$Department of Physics, The College of New Jersey, P.O.\ Box 7718, Ewing, NJ 08628-0718, USA \\
$^{4}$Abastumani Observatory, Mt. Kanobili, 0301 Abastumani, Georgia \\
$^{5}$Engelhardt Astronomical Observatory, Kazan Federal University, Tatarstan, Russia \\
$^{6}$Center for Astrophysics, Guangzhou University, Guangzhou, 510006, China \\
$^{7}$Landessternwarte, Zentrum für Astronomie der Universität Heidelberg, K$\ddot{o}$nigstuhl 12, 69117 Heidelberg, Germany \\
$^{8}$American Association of Variable Star Observers (AAVSO), 49 Bay State Rd., Cambridge, MA 02138, USA \\
$^{9}$Group M1, Centro Astron$\acute{o}$mico de Avila, Madrid, Spain \\
$^{10}$Astronomical Observatory, Volgina 7, 11060 Belgrade, Serbia \\
$^{11}$Department of Physics and Astronomy, Faculty of Natural Sciences, University of Shumen, \\
~~115, Universitetska Str., 9700 Shumen, Bulgaria \\
}

\begin{document}

\date{Accepted ....... Received  ......; in original form ......}

\pagerange{\pageref{firstpage}--\pageref{lastpage}} \pubyear{2010}

\maketitle

\label{firstpage}
\begin{abstract}

We present the results of photometric observations of three TeV blazars, 3C 66A, S5 0954+658 and BL Lacertae, during the period
2013--2017. Our extensive observations were performed in a total of 360 nights which produced $\sim$6820 image frames in BVRI bands.
We study flux and spectral variability of these blazars on these lengthy timescales. We also examine the optical Spectral Energy 
Distributions of these blazars, which are crucial in understanding the emission mechanism of long-term variability in blazars. 
All three TeV blazars exhibited strong flux variability during our observations. The colour variations are mildly 
chromatic on long timescales for two of them. The nature of the long-term variability of 3C 66A and S5 0954$+$658 is consistent with a model of a non-thermal
variable component that has a continuous injection of relativistic electrons with power law distributions around 4.3 and 4.6, respectively.
However, the long-term flux and colour variability of BL Lac suggests that these can arise from modest changes in velocities or
 viewing angle toward the emission region, leading to variations in the Doppler boosting of the radiation by a factor $\sim1.2$ over the period of these observations.

\end{abstract}

\begin{keywords}
galaxies: active -- BL Lacertae objects: general -- quasars:  individual (3C 66A, S5 9854$+$658, BL Lac)  -- galaxies: photometry
\end{keywords}

\section{Introduction}

Blazars are well known for their rapid flux variability across the whole electromagnetic spectrum, strong optical
and radio polarization and non-thermal anisotropic emission (Ulrich, Maraschi \& Urry 1997). The dominant radiation
properties of blazars are interpreted by the relativistically beaming effects which are believed to originate
from strongly Doppler boosted relativistic jets that are being viewed at angle $\leq 10^{\circ}$ with respect to
the line of sight (Urry \& Padovani 1995). Flat spectrum radio quasars along with the BL Lacertae objects are now normally said to
constitute the
blazar class of active galactic nuclei (AGN).  The Doppler beaming effects of blazar jets viewed at small angles can strongly amplify the perceived 
luminosity and simultaneously shorten their apparent time-scales; this property
 makes blazars ideal targets to use  flux variability studies to probe their central engines.

The broad band spectral energy distribution (SEDs) of blazars are characterized by  double peaked structures. The low energy
peak is dominated by  the synchrotron radiation from relativistic electrons. In the leptonic
scenarios, the high energy peak can be due to the inverse Compton scattering of lower energy synchrotron photons 
from the same electron population in synchrotron-self-Compton (SSC scenario; e.g. Kirk, Rieger \& Mastichiadis 1998) 
or of external photons from broad line region, accretion disc or dusty torus in an external Compton (EC) scenario (e.g., 
Sikora, Begelman \& Rees 1994). In the hadronic scenarios, it could be due to synchrotron emission from protons or, more likely 
from secondary decay products of charged pions (e.g. Atoyan \& Dermer 2003; B{\"o}ttcher et al.\ 2013).
Based on the location of the synchrotron peak, blazars are classified into low, intermediate and high energy peaked blazars
 (LBL, IBL and HBL, respectively; Padovani \& Giommi 1995). Abdo et al. (2010) classified blazars based
on the location of the synchrotron peak frequency, $\nu_{s}$. If $\nu_{s}$ $\le$ $10^{14}$ Hz (in the infrared), it is classified
as a low spectral peak (LSP) source; if it is in the optical--ultraviolet range ($10^{14}$ $\le$ $\nu_{s}$ $\le$ $10^{15}$ Hz),
it is classified as an intermediate spectral peak (ISP) source, and if it lies in the X-ray regime ($\nu_{s}$ $\ge$ $10^{15}$ Hz),
it is classified as a high spectral peak (HSP) source.

Blazars are variable on diverse time-scales ranging from minutes through months to even decades. Their variability timescales
can be broadly divided into three classes. Intra-day variability (IDV) timescales range from a few minutes to several
hours and flux changes by a few tenths of magnitude (e.g. Wagner \& Witzel 1995).
Short term variability (STV) and Long Term variability (LTV) have timescales ranging from several days to months
and months to decades, respectively (Fan et al.\ 2009; Gaur et al.\ 2015a, b; Gupta et al.\ 2016).

IDV in blazars almost certainly arises due to purely intrinsic phenomenon such as the interaction of shocks with small scale particle or 
magnetic field irregularities present in the jet (e.g.\ Marscher 2014; Calafut \& Wiita 2015) or the production of 
ultrarelativistic mini-jets within the jet (e.g. Giannios et al.\ 2009). When the viewing angle to a moving, discrete emitting 
region changes, it causes
variable Doppler boosting of the emitting radiation (e.g.\ Larionov et al.\ 2010; Raiteri et al.\ 2013).  LTV is usually attributed to a
mixture of intrinsic and extrinsic mechanisms. Here we consider intrinsic mechanisms to  involve shocks propagating down  twisted jets or  plasma blobs 
moving through some helical structure in the magnetized jets (e.g.\ Marscher et al. 2008). Extrinsic mechanisms would involve the
geometrical effects that result in an overall bending of the jets, either through instabilities (e.g.\  Pollack et al.\ 2016), or through orbital
motion (e.g., Valtonen \& Wiik 2012).  

Photometry is a powerful tool to study the structure and radiation mechanism of blazars by measuring their
variability timescales, amplitude and duty cycle.
In this work, we performed extensive optical observations of three TeV blazars,  3C 66A, S5 0954$+$658 and BL Lac to study
their flux and colour variability on long-term time scales. We also studied the Spectral Energy Distributions (SEDs) of these blazars
across optical bands. Our photometric observations were made at several telescopes during the period 2013--2017. 
A total of 360 nights of observations are reported here allowing us to search
for and characterize their variability on rather long timescales. This study is a follow up of our earlier extensive optical
 monitoring of TeV blazars covering the period 2010--2012 (Gaur et al. 2015a,b; Gupta et al. 2016). Two of these sources, S5 0954
$+$658 and BL Lac were also observed for IDV on many nights and those data have been studied in Bachev (2015) and Gaur et al.\ (2017), respectively.
 In Section 2 we briefly describe our observations and data reductions. We present our results in Section 3. Sections 4 gives a discussion and conclusions.

\section{Observations and Data Reduction}

Our observations of three TeV blazars, 3C 66A, S5 0954$+$658 and BL Lacertae began in January 2013 and concluded in June 2017. 
The logs of observations for these blazars are provided individually in Table 1. 
 Complete observation log of these three sources are provided
in Supplementary Material. Our observations 
were made using 
seven telescopes in Bulgaria, Georgia, Greece, Serbia and Spain. For 3C 66A, we also used archival data of the Steward Observatory 
(Smith 2009) in V band for a total of 165 nights.                             
Details about the telescopes in Bulgaria, Greece and Georgia are given in Gaur et al.\ (2012, 2015a); details about Serbian Telescope
 are described in Gupta et al.\ (2016, Table 1) and details about the telescope in Spain are provided in Gupta et al.\ (2016).
The standard data reduction methods we employed on all observations are described in
detail in Section 3 of Gaur et al.\ (2012). Typical seeing varied between 1--3 arcsec during these measurements.

To summarize,  standard data reduction was performed using \texttt{IRAF}\footnote{IRAF is distributed by the National Optical Astronomy
Observatories, which are operated by the Association of Universities for Research in Astronomy, Inc., under cooperative agreement with the
National Science Foundation.}, including bias subtraction and flat-field division. Instrumental magnitudes of the source
 as well as comparison stars in its field (Villata et al.\ 1998) were produced
using the \texttt{IRAF} package \texttt{DAOPHOT}\footnote{Dominion Astrophysical
Observatory Photometry software}; normally an aperture radius of $6^{\arcsec}$ was used along with
a sky annulus of 7.5--10 arcsec.

Data from the different telescopes are collected as instrumental magnitudes of the blazar and reference stars
so that we can apply the same analysis and calibration procedures to all the datasets. Instrumental magnitudes
are obtained by the above mentioned standard aperture photometry procedures by the observers. Data from one 
observatory (35.6 cm Telescope, Spain) was provided in calibrated magnitudes. During the compilations of light curves, we discarded a small number of clearly bad and unreliable points. 

\begin{table}
\noindent
{\caption {Observation log of optical photometric observations of the blazar 3C 66A.} }
\centering
\begin{tabular}{lcc} \hline
~~~~Date & Telescope  & Data Points \\
dd mm yyyy   &     &B, V, R, I \\
\hline
11.01.2013  &  H  &0,0,1,0 \\
12.01.2013  &  H  &0,0,1,0 \\
15.01.2013  &  F  &0,0,4,0 \\
16.01.2013  &  H  &0,0,1,0 \\
\hline
\end{tabular}

\vspace*{0.05in}
\noindent
(This table is available in its entirety in a machine-readable form in the online journal. A portion is
shown here for guidance regarding its form and content)

\footnotesize
A: 50/70-cm Schmidt Telescope at  National Astronomical Observatory Observatory, Rozhen, Bulgaria   \\
B: 1.3-m Skinakas Observatory, Crete, Greece \\
C: 2m RCC, National Astronomical Observatory, Rozhen, Bulgaria   \\
D: 60-cm Cassegrain Telescope at Astronomical Observatory Belogradchik, Bulgaria \\
E: 60-cm Cassegrain Telescope, Astronomical Station Vidojevica (ASV), Serbia \\
F: 70-cm meniscus telescope at Abastumani Observatory, Georgia  \\
G: 35.6 cm Telescope at Observatorio Astronomico Las Casqueras, Spain   \\
H: 2.3-m Bok Telescope and 1.54-m Kuiper Telescope at Steward Observatory, Arizona, USA   \\

\end{table}

\section{Results}

\subsection{3C 66A}

3C 66A is one of the most luminous blazars at TeV $\gamma$-rays (Very High Energy (VHE) , E $>$ 100 GeV). It was detected at VHE
$\gamma$-rays by VERITAS (Acciari et al.\ 2009). It is classified as a BL Lac object (Maccagni et al.\ 1987). Recently, it has been
proposed that this blazar is hosted in a galaxy belonging to a cluster at $z=0.340$ (Torres-Zafra et al.\ 2016). The synchrotron peak of
3C 66A is located between $10^{15}$ and $10^{16}$ Hz and hence it is best classified as an intermediate  frequency peaked BL Lac object 
(Perri et al.\ 2003; Fan et al.\ 2016).

3C 66A was the target of the multi-wavelength campaign carried out by the Whole Earth Blazar Telescope (WEBT) during its optically bright state in 2003--2004,
involving observations at radio, infrared, optical, X-ray and $\gamma$-ray bands  (B{\"o}ttcher et al.\ 2005). At optical frequencies, 
it showed several flaring events on  time-scales of days and flux changes of $\sim$5\% on timescales of hours.
Again, WEBT organized a multi-wavelength campaign in 2007--2008 in the high optical state of the source and it exhibited several
bright flares on time-scales of days and even on hours (B{\"o}ttcher et al.\ 2007). This campaign caught an exceptional outburst
around September 2007 reaching peak brightness at R $\sim$13.4, with a lack of spectral variability in this high state.
 Abdo et al.\ (2011) also used multi-wavelength observations of this source to study its behaviour
in flaring and quiescent states. Recently, Kaur et al.\ (2017) studied the optical IDV of 3C 66A over more than 10 years
(from 2005--2016) and found significant variations in the optical flux on long-term time-scales with a large number of flares
superimposed on the slowly varying pattern.

\subsubsection{Long term flux and colour variability}

The LTV light curve (LC) of 3C 66A is shown in Fig. 1 for a total of 295 nights over nearly four years from 
January 2013--January 2017.  The B, V, R and I LCs are shown from top to bottom panels. 
We observed this source on 130 nights for a total of 697 photometric frames with maximum sampling in V and R bands.
We averaged the magnitudes on the daily basis to understand the long-term behaviour of the source.
We also used the archival data of the Steward Observatory (Smith 2009) available online in V band for 165 nights. 
In order to visualize the nature of variability more clearly, we plot the R band LC in the upper panel of Fig.\ 2. It can be 
seen that the R band LC shows many small sub-flares superimposed on the long-term trends. First, there is a decaying trend in 
the LC from JD 6400--6830 where the magnitude drops by 0.62. There is one major outburst during JD 6830--7370, peaking at 
around JD=7200 with R=14.07. The source magnitude changes by 0.64 from $m_{R}$=14.7 to 14.1 and back again during this outburst 
over about 370 days. After that, there is again an overall rising trend during JD 7370--7800 by $\Delta$m=0.53.

 To set some context, 3C 66A exhibited an exceptional outburst around September 2007, reaching peak brightness at $\sim$13.4 (B{\"o}ttcher et al.\ 2009).
 Also, Kaur et al.\ (2017) presented a decade long LC of 3C 66A and found this source reaching a peak brightness 
of 13.40 in R band on 1st November 2010 with a minimum magnitude of only 15.1 in August 2011. 
During our observing campaign, the average R band magnitude is 14.4 with a maximum of 14.07 magnitude on JD=7200
and a minimum of 14.72 magnitude on JD=7340. Hence, we can say that the source is mostly in a relatively high state during 
our observing campaign. LCs in other bands also shows similar behaviour. The mean magnitudes
in the B, V and I bands are 15.43, 14.80 and 13.83, respectively. The maximum and minimum brightness magnitudes are: in B band, max=15.07, min=15.80; V band, max=14.36, min=15.23; and I band,  max=13.50, min=14.16. While Fig.\ 1 makes it clear that we have many fewer
data points in the B and I bands, fortuitously, several were taken close to both the minima and greatest maxima.

\begin{table}
{\caption {Spectral indices of 3C 66A and S5 0954+658 for various flux states.} }
\noindent
\begin{tabular}{ccc} \hline 
                   \hline
Source  &Flux        &Spectral \\
        &(mJy)            &Index \\ \hline

 3C 66A &7.75            &1.64   \\
&4.76            &1.66   \\
&3.68            &1.68    \\ \hline
 S5 0954+658 &18.01           &1.84  \\
&4.31            &1.90   \\
&0.34            &2.72   \\\hline

\end{tabular}
\end{table}

\begin{table}
{\caption {Properties of variable emission components of 3C 66A and S5 0954+658} }
\noindent
\begin{tabular}{cccc} \hline 
                   \hline
Source       &Band        &Number of       &log (m) \\
             &            &Observations  & \\ \hline
3C66A        &I           &65  &$+$0.114   \\
             &R           &--   &  \\
             &V           &66  &$-$0.081  \\
             &B           &65  & $-$0.265  \\ \hline
S5 0954+658  &I           &43  &$+$0.110  \\
             &R           &--   &--  \\
             &V           &26  &$-$0.096\\
             &B           &38  &$-$0.320   \\  \hline

\end{tabular}
\end{table}

\begin{figure*}
\centering
\includegraphics[width=12cm , angle=0]{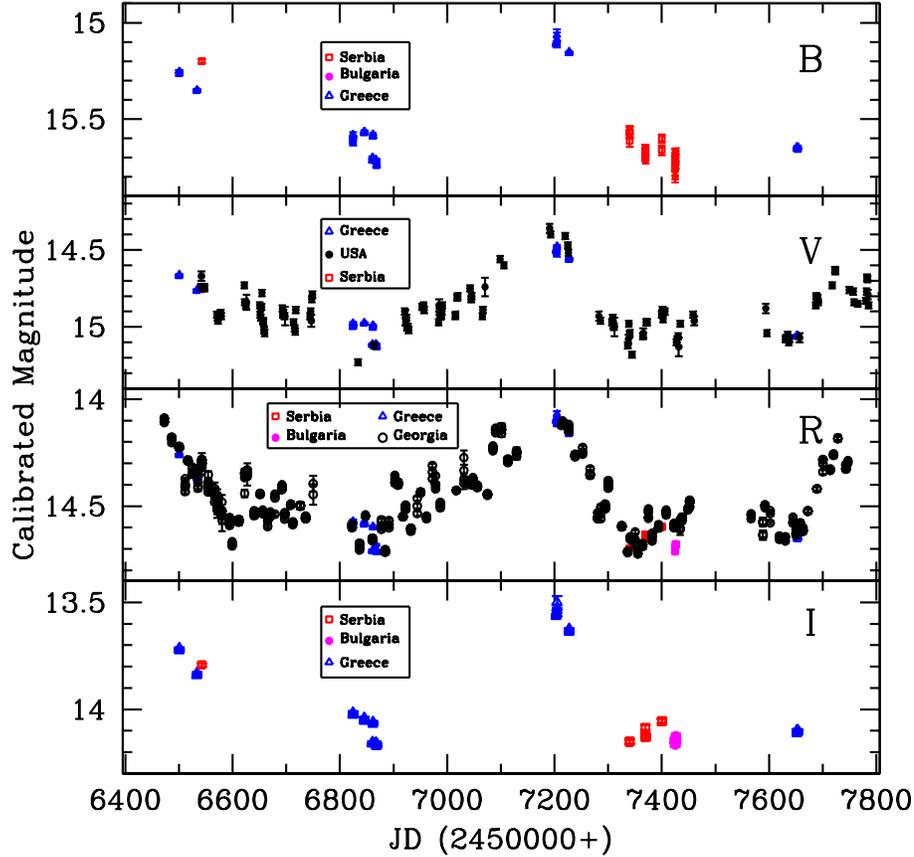}
\caption{Long term light curve of 3C 66A during the period 2013--2017.  The top to bottom panels respectively show B, V, R and I calibrated magnitudes and different symbols indicate data from different telescopes as defined in 
each peanel separately.}
 \end{figure*}

\begin{figure*}
\centering
\includegraphics[width=12cm , angle=0]{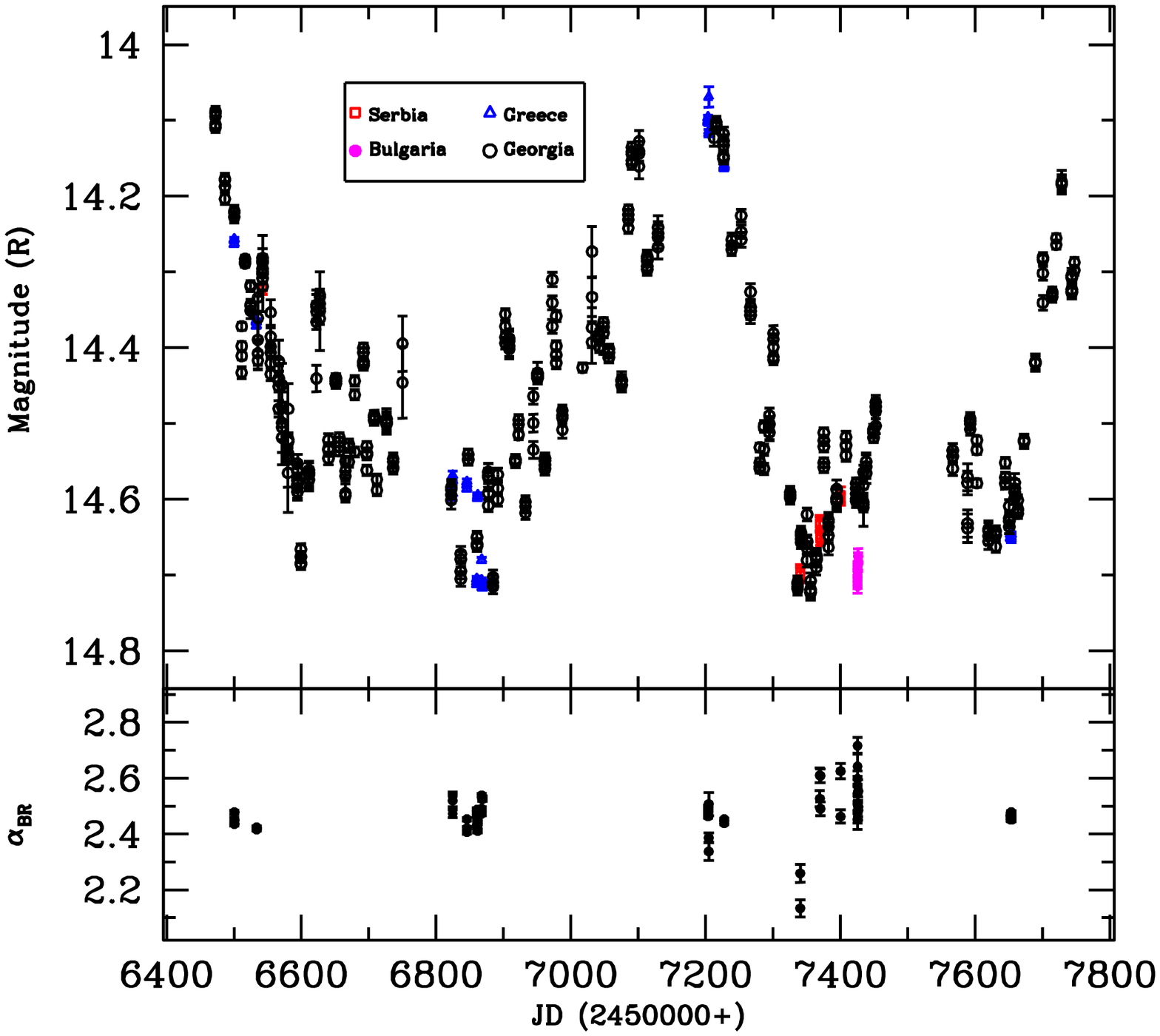}
\caption{R band LC of 3C 66A (upper panel) and the variation of spectral indices $\alpha_{BR}$ with respect to JD (lower panel). 
Spectral indices are calculated for those nights where we have quasi-simultaneous data in
B and R bands. }
 \end{figure*}

We estimated the average spectral indices as calculated in Wierzcholska et al.\ (2015):
\begin{equation}
\alpha_{BR}= \frac {0.4 (B-R)}{log(\nu_{B}/\nu_{R})},
\end{equation}
where $\nu_{B}$  and $\nu_{R}$ are the effective frequencies of the respective bands.
We calculated the spectral index for only those nights of our observation where we have simultaneous data in B and R bands  
and they are shown in the lower panel of Fig.\ 2. 
The $\alpha_{BR}$ slope is very steep during the observing period and varies between 2.1--2.7. This plot again indicates 
that the emission is dominated by synchrotron emission.

We searched for a relationship between colour-index with respect to magnitude which is needed to ascertain whether 
flux changes are associated with the spectral changes. We used the simultaneous observed data in B, V and R bands for a 
particular night. Figure 3 shows the color variability diagram during the period 2013--2017. The color bar indicates the progression of time in 
JD (2450000+). We have 
fitted straight lines (CI = $m \times$R+$c$) to colour index, CI, versus R magnitude plot. 
We found $m$=0.021$\pm$0.019; $c$=1.30$\pm$0.29, with linear Pearson correlation coefficient $r$=0.13 indicating 
 $p$=0.069.
Here  a positive correlation is defined as a positive slope between the colour index and magnitude of the blazar which
implies that the source tends to be bluer when it brightens or redder when it dims. An opposite correlation
with negative slope implies the source follows a redder when brighter behaviour.
 As the nominal values of the slopes are small and do not exceed twice the errors, we did not find any significant correlation 
between magnitude and colour index for this blazar. In the previous studies
 of spectral variability, B{\"o}ttcher et al.\ (2005) found a weak positive 
correlation between (B$-$R) versus R magnitude when 
the source was in a low brightness state with R $\geq$14.0 but no clear correlation was found in brighter states of the source. 
Also, this source was a target of WEBT campaign again in 2007--2008 when it was in a high optical state, reaching a peak brightness
of R $\sim$13.4 but then it did not display spectral variability (B{\"o}ttcher et al.\ 2009).

\begin{figure}
\centering
\includegraphics[width=8cm , angle=0]{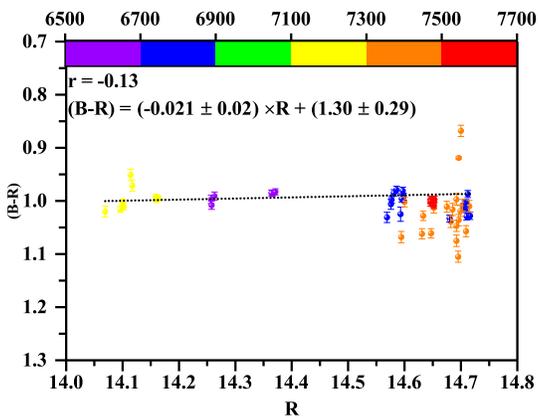}
\caption{Colour versus magnitude relation for 3C 66A}
 \end{figure}

\subsubsection{Spectral Energy Distribution (SED)}

To learn about the origin of variability on longer timescales, we studied the SED in optical bands in different 
flux states. We adopted the assumption by Hagen-Thorn et al.\ 2004 (and considered later by Larionov et al.\ 2008; 2010) that the 
flux changes over the relevant set of observations arise from a single variable source (or multiple sources with similar SED).
 Hence, if the relative SEDs do not change and the variability is caused only by flux variations in that source region, then if $n$ is the number  of  bands in which observations are made, the measured points should lie on straight lines in the
{\it n}-dimenssional flux density space.  As noted by Larionov et al.\ (2010), the converse would also be true, so that an observed linear 
relation between  flux densities 
at two different wavelengths while the brightness varied means that the flux ratio, or slope is invariant.  If linear relations were detected between multiple bands that would mean that the slopes of those lines could yield the constant relative SED of that variable source component.

\begin{figure*}
\centering
\includegraphics[width=5cm , angle=0]{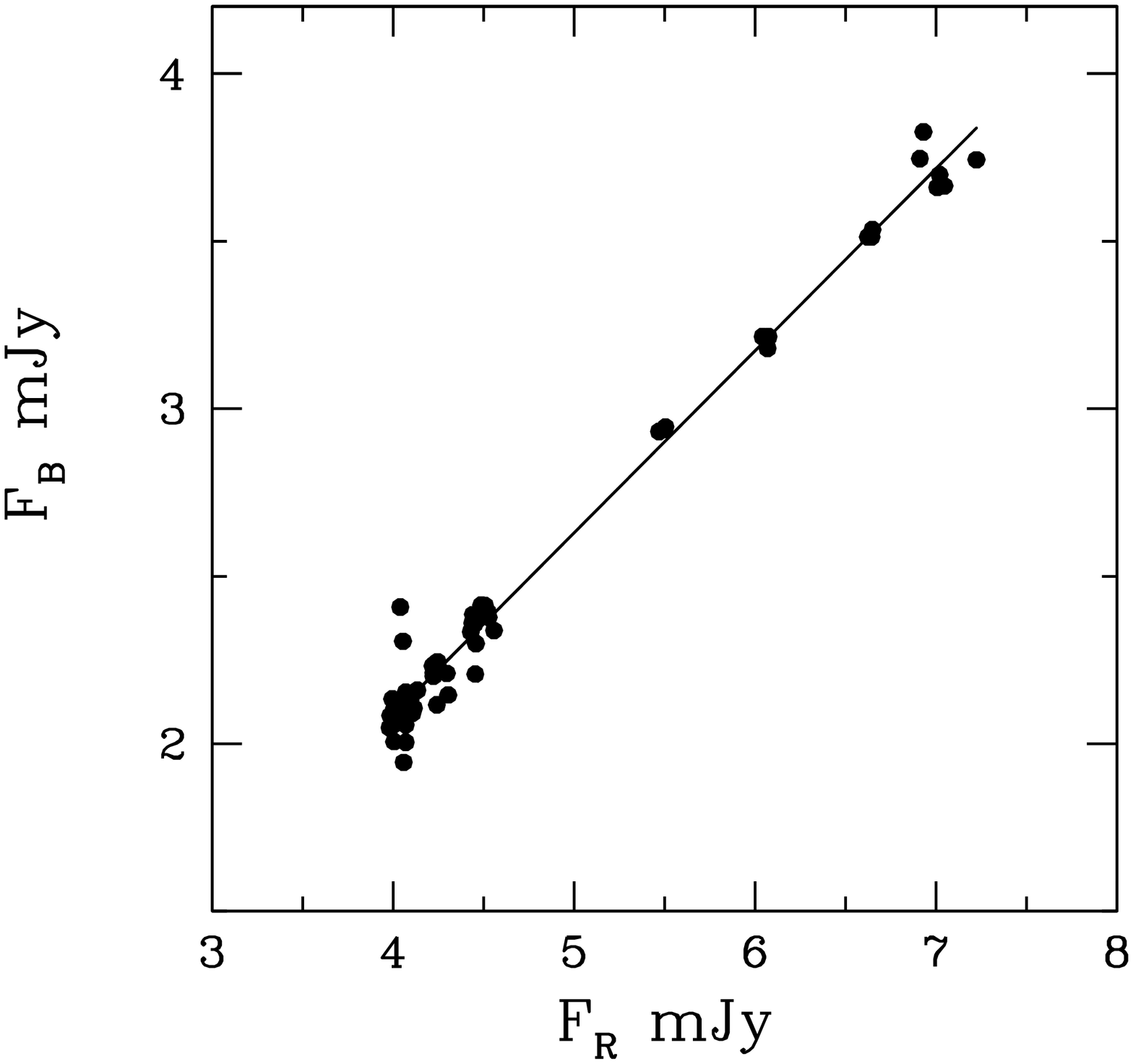}
\includegraphics[width=5cm , angle=0]{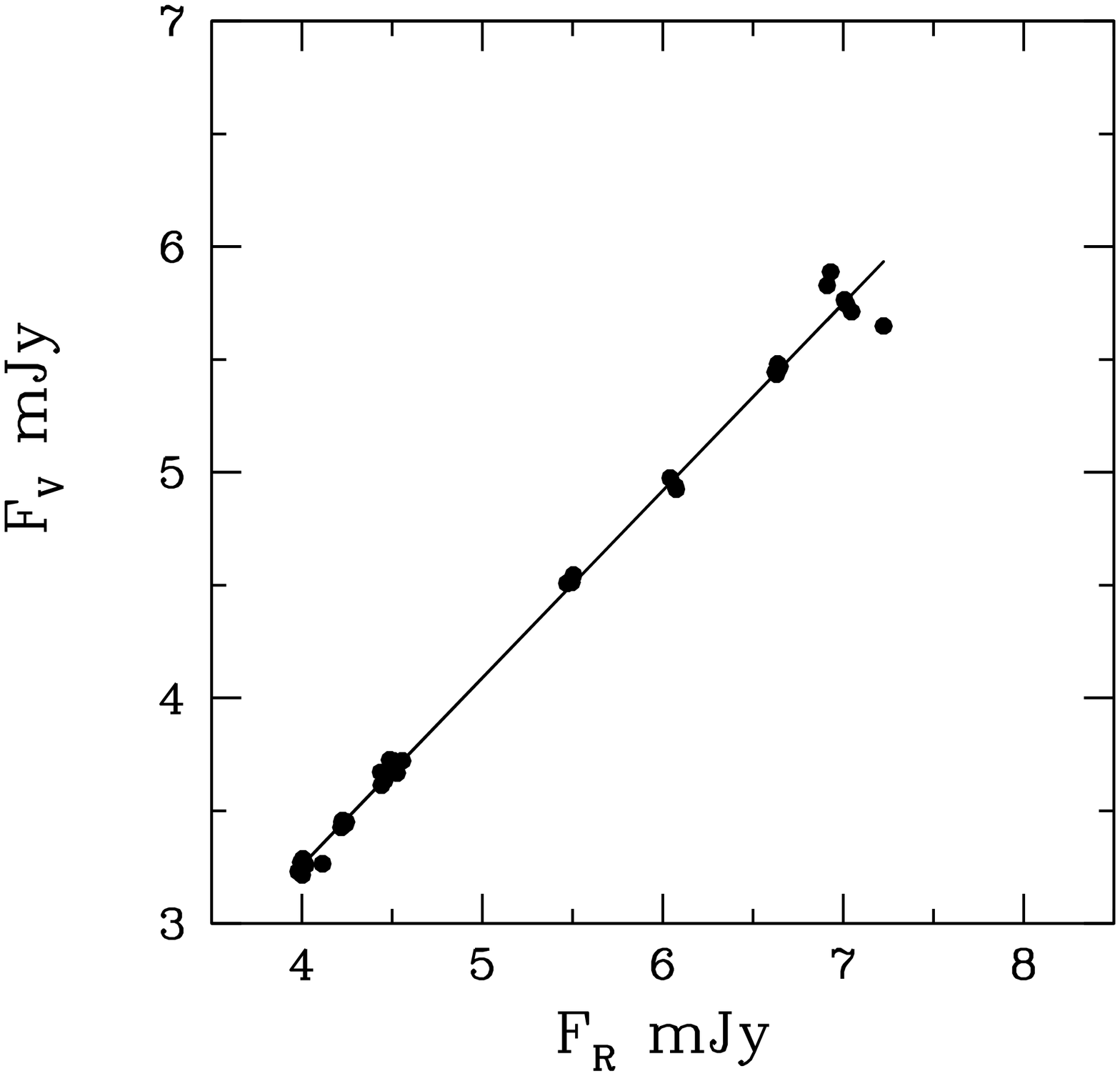}
\includegraphics[width=5cm , angle=0]{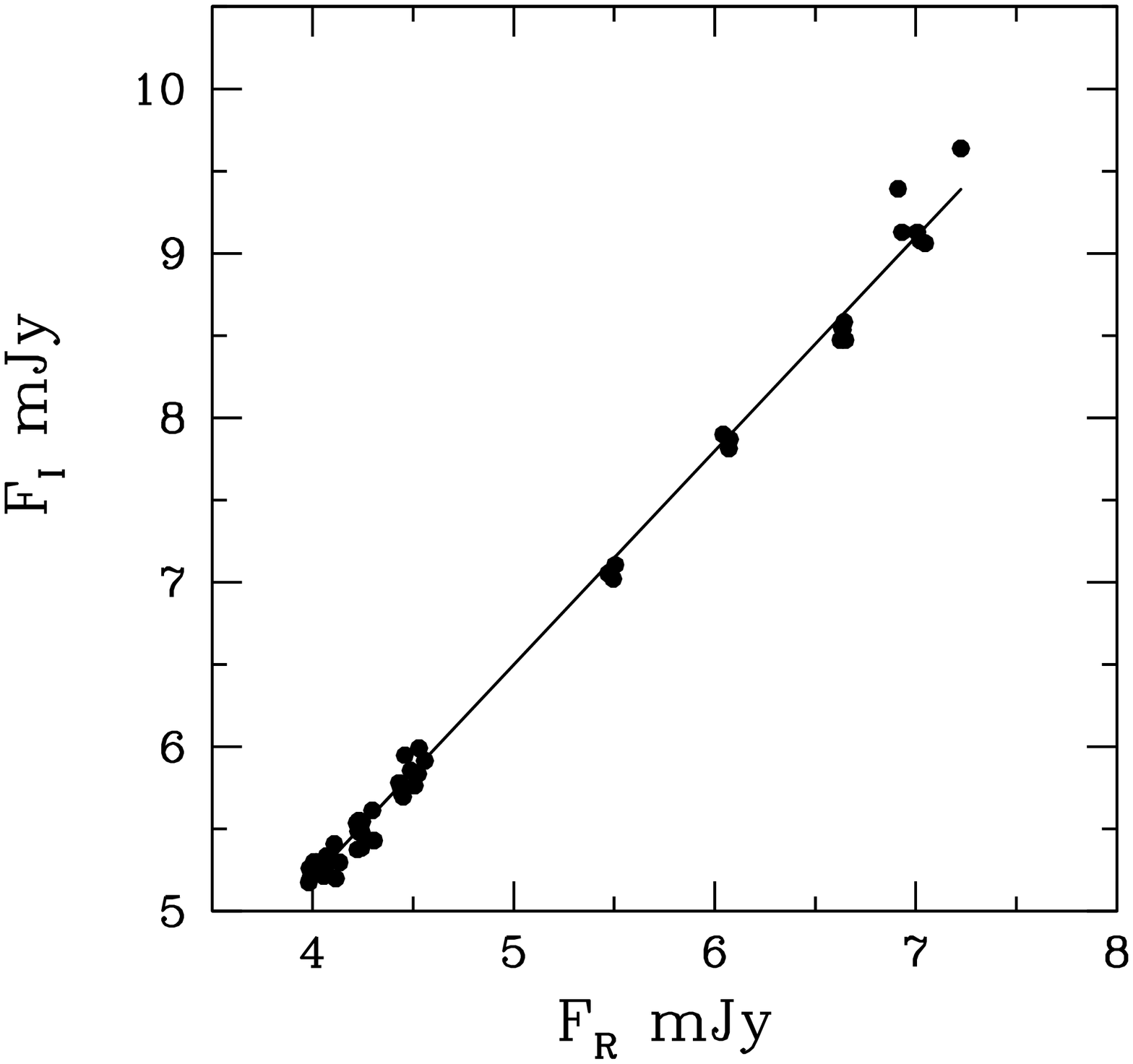}
\caption{Flux--flux dependence of 3C 66A, where the B (left), V (center)  and I (right) band fluxes are  plotted against the R flux; 
 lines represent linear regressions.}
 \end{figure*}


\begin{figure}
\centering
\includegraphics[width=8cm , angle=0]{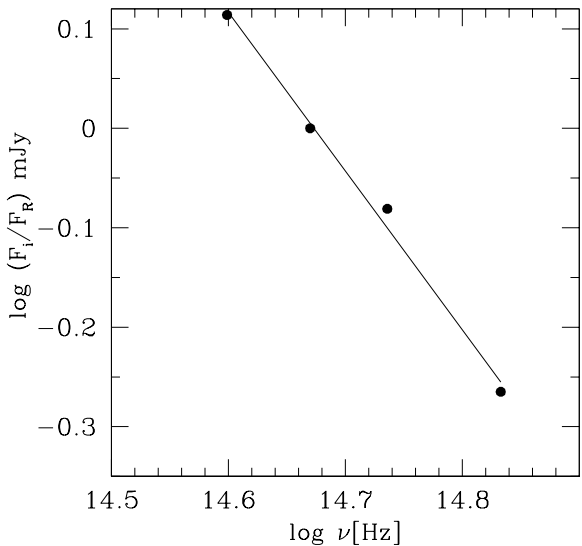}
\caption{Relative SED of the variable component of 3C 66A, normalized with respect to R band.}
 \end{figure}

Hence we searched for the flux--flux relations between the B, V, R and I bands. First, we transformed magnitudes into
de-reddened flux densities using the Galactic absorption from Schlegel, Finkbeiner \& Davis (1998); see Bessell et al.\ (1998).
The flux--flux relations are shown in Fig.\ 4. it can be seen from the figure that flux ratios follow linear
dependencies, and we used R band data to obtain relations as: $F_{i} = c_{i} + m_{i} ~ F_{R}$ where $i=$ B, V and I bands. These
linear regressions are used to construct the SED during different flux states of the source observed during our campaign i.e., maximum (7.75 mJy);
average (4.76 mJy) and minimum (3.68 mJy) fluxes.
All the three curves are fitted with a power law, i.e.,  $F_{\nu} \propto \nu^{-\alpha}$, for the spectral index $\alpha$. Spectral indices for all the mentioned
 flux states are provided in the upper part of Table 2 and are constant at $\simeq 1.66$.
 As mentioned above, slopes of the linear regressions can also be used to construct the
relative SED of the variable source.
 The values of these slopes and the number of observational points used in producing them are provided in Table 3. A SED of 3C 66A which is normalized
to the R band is shown in Fig.\ 5. We derive a power law slope of the variable component to be $\alpha$=1.60$\pm$0.09.
As long as we can assume that additional components are responsible for the observed variability and the relative SED
of this variable portion is not varying,
 it can be concluded that the origin of variability of this blazar is intrinsic. It then can be accommodated within a non-thermal
variable component which has a continuous injection of relativistic electrons with a power law energy distribution of $p \simeq$4.3
as $\alpha \simeq (p-1)/2$.

\subsection{S5 0954+658}

S5 0954+658 is a BL Lac object with a claimed redshift of $z=0.367$ (Stickel et al.\ 1993); however, Landoni et al.\ (2015) found a 
featureless spectrum and a limit of $z > 0.45$, which is based on non-detection of the host galaxy.
This source is optically active and its optical variability was first studied by Wagner et  al.\ (1993).
Raiteri et al.\ (1999) studied a four year long 
optical light curve and  detected fast large amplitude variations. They also found that the variations over long time-scales
for this source were not associated with spectral variations. Papadakis et al.\ (2004) studied intra-day flux and spectral 
variability of this source during 2001--2002. Hagen-Thorn et al.\ (2015) studied the optical flux and polarization 
variability of this blazar during 2008--2012. They found flux variations exceeded 2.5 mag 
and a degree of polarization that reached
40\%. Larionov et al.\ (2011) also performed photopolarimetric observations of S5 0954+658 during 2011 and found 
it to have been in a faint state in mid-January 2011 with a brightness level R $\sim$17.6, but by the middle of March the optical brightnes
rose up to $\sim$14.8 mag, showing flare-like behaviour. During the rise of the flux, the position
angle of the optical polarization rotated smoothly over more than 200 degrees.  
Bachev (2015) presented the intra-night monitoring of the blazar during February 2015 and found violent 
variations on very short timescales of $\sim 0.1$ mag within $\sim10$ min.

\begin{figure*}
\centering
\includegraphics[width=12cm , angle=0]{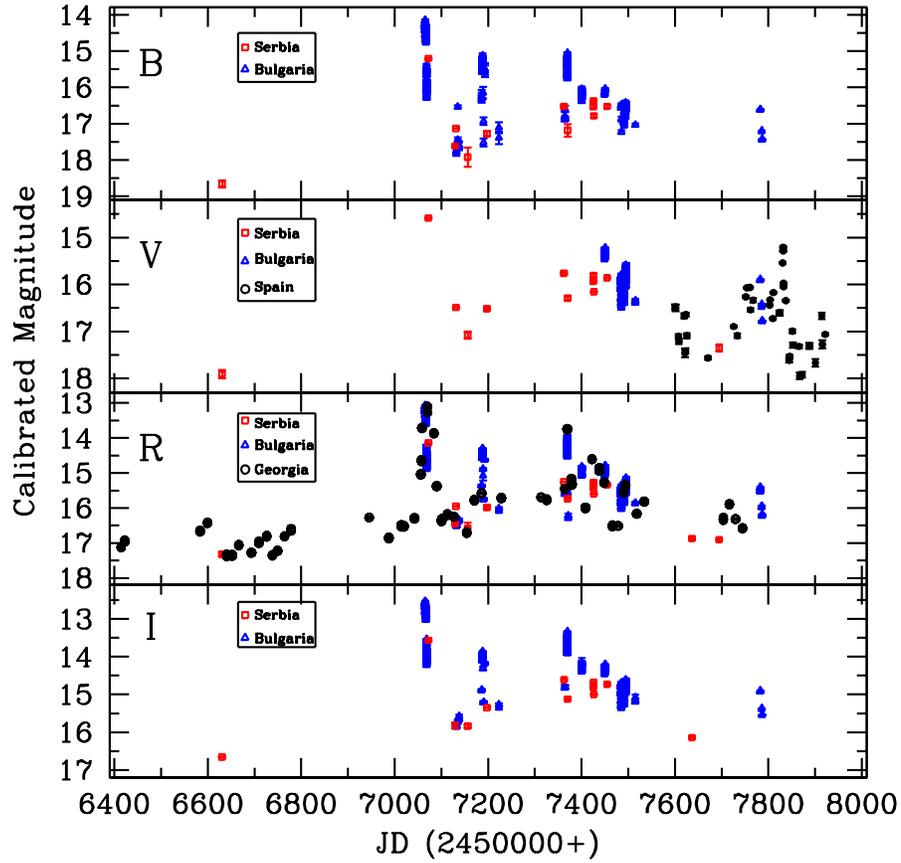}
\caption{Long term LC of S5 0954+658 during the period 2013--2017, with symbols as in Fig.\ 1.}
 \end{figure*}

\begin{figure*}
\centering
\includegraphics[width=9cm , angle=0]{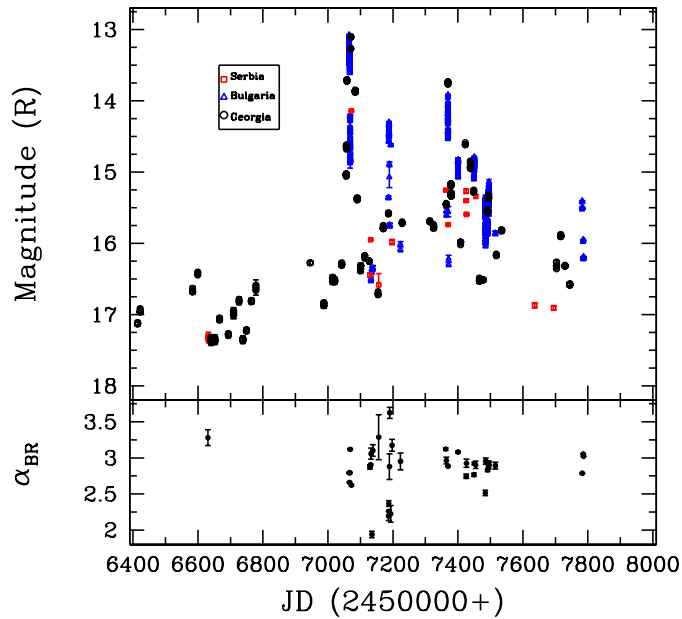}
\caption{R band LC of S4 0954+658 (upper panel) and the variation of spectral indices $\alpha_{BR}$ (lower panel) with respect to JD.
(lower panel).}
 \end{figure*}

\subsubsection{Long term flux and colour variability}

The LTV LC of S5 0954$+$658 is shown in Fig.\ 6 for a total of 170 nights over nearly four years from
February 2013--June 2017. B, V, R and I LCs are respectively shown from the top to the bottom panels of the above figure.
To understand the long-term behaviour of this source it was observed for a total of 1756 photometric frames with maximum sampling 
in V and R bands. In order to visualize the nature of variability more clearly, we replot the R band LC on an expanded scale in 
upper panel of Fig.\ 7; it displays many short subflares superimposed on the longer term trends. 
The source was in a low optical state during the period MJD 6400--7000, with the mean magnitude of this source
during this period $\sim$17. After this period, the source flux  increased significantly and reached up to an
R magnitude of $\sim13$ as was also reported by Bachev (2015). Until MJD=7600 this source displayed many flare-like events.
There also was a strong flare peaking around MJD=7810 which was only caught in V band measurements; it involved a rise of $\sim$2 
mag over $\sim$ 100 days and a more rapid decline. During our observing campaign, the faintest R band magnitude was 17.40 at 
MJD=6640 and the brightest was 13.19 magnitude on MJD=7070. LCs in other bands also show similar behaviours. The faintest and 
brightest magnitudes, respectively in B band were 18.65 and 14.15; in V band, 17.91 and 15.31; and in I band, 16.66 and 12.64.  
Hence, during our  campaign we observed the source in its low state as well as in its high state.

In order to search for the relationship between colour index with respect to the magnitude, we used the simultaneous data
in B and R bands. A color variability diagram is shown in Figure 8 where the color bar indicates the progression of time in JD(2450000+). 
The straight line fitted to colour index versus R magnitude gives $m$=0.048$\pm$0.010; $c$=0.45$\pm$0.21=0.337; $p$=0.51 which indicates a 
positive correlation between colour and magnitude.
Raiteri et al. (1999) studied the long term light curve of this source and detected large amplitude variations but they were
not associated with the spectral variations.

\begin{figure}
\centering
\includegraphics[width=8cm , angle=0]{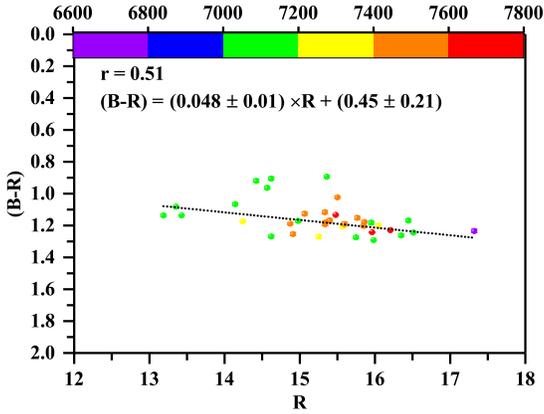}
\caption{Colour versus magnitude relation of S4 0954+658}
 \end{figure}

\subsubsection{Spectral Energy Distribution}

As we did for 3C 66A, we examined the flux--flux relations of S5 0954$+$658, shown in Fig.\ 9 and found that they follow linear regressions.  Using the slopes
of these linear regressions, we constructed SEDs at maximum (18.01 mJy); average (4.31 mJy) and minimum (0.34 mJy) levels, respectively.
 We fitted these SEDs using $F_{\nu}$ $\propto$ $\nu^{\alpha}$ and their respective slopes are provided in 
Table 2.  These indicate that the slope is  flatter when the flux increases, which is a common
property of High Synchrotron Peaked blazars (e.g., Vagnetti et al.\ 2003; Hu et al.\ 
2006; Gaur et al.\ 2012). According to Hagen-Thorn et al.\ (2004), if such flux--flux relations are linear, a variable emission
component is likely to be responsible for the variability. Hence, we constructed the relative SED of the variable emission in Fig.\ 10
using the slopes of flux--flux relations provided in Table 3.   These indicate that the slope of the variable component is 1.82$\pm$0.15. 
Hence, the variability of S5 0954+658 can be interpreted in terms of a non-thermal variable component 
that has a continuous injection of relativistic electrons with a steep power-law energy distribution of $p \sim$4.64.

\begin{figure*}
\centering
\includegraphics[width=5cm , angle=0]{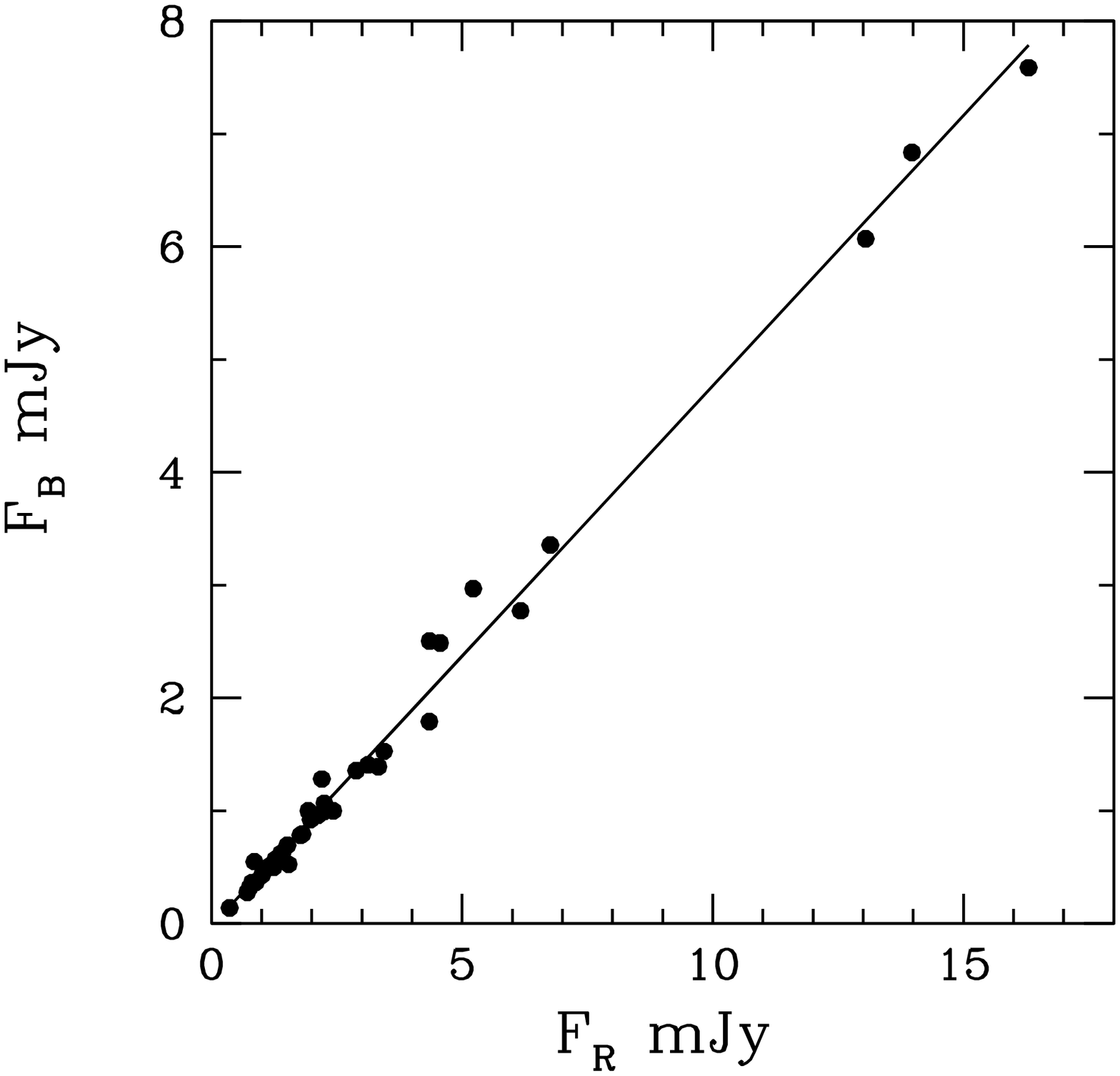}
\includegraphics[width=5cm , angle=0]{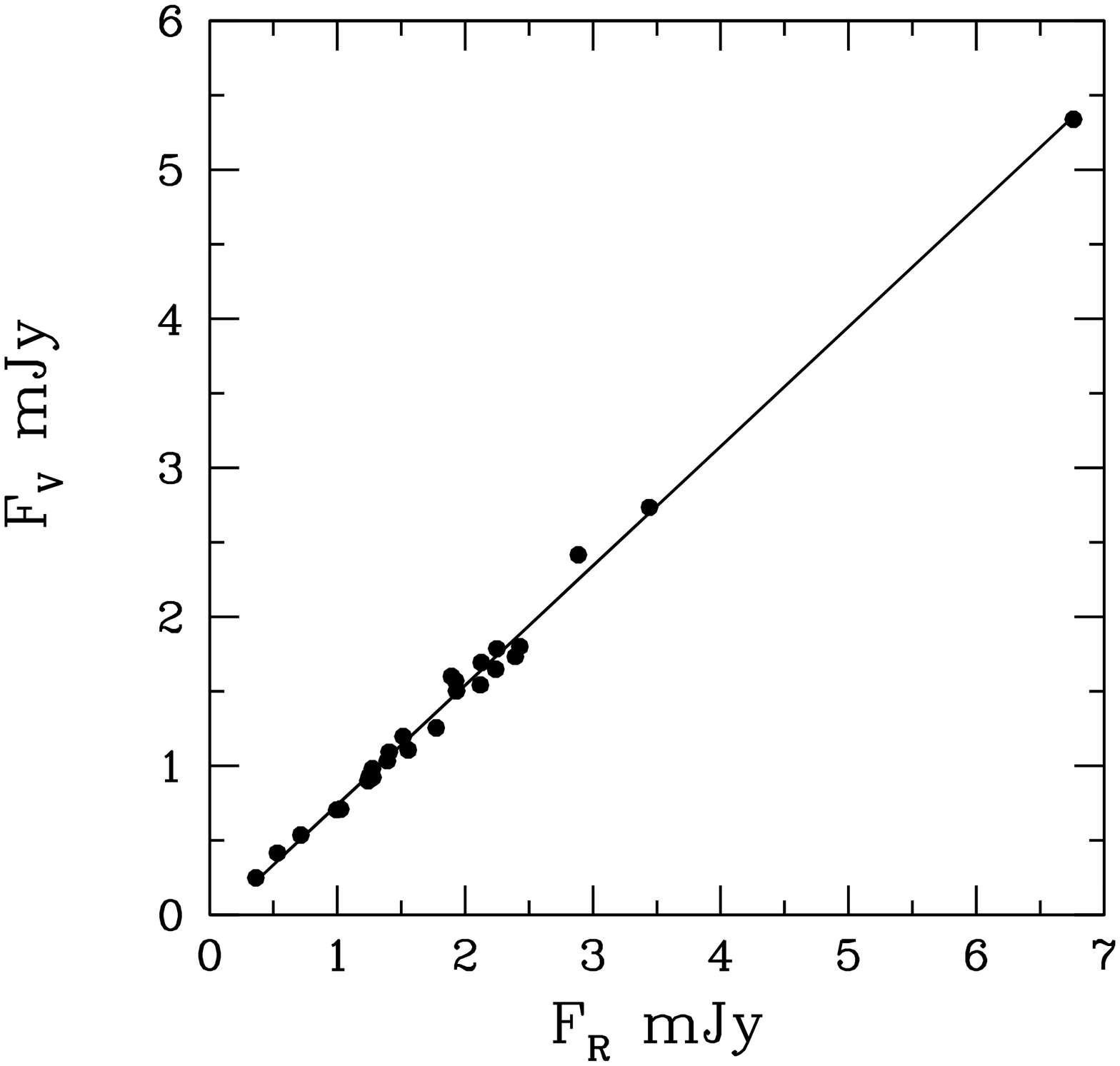}
\includegraphics[width=5cm , angle=0]{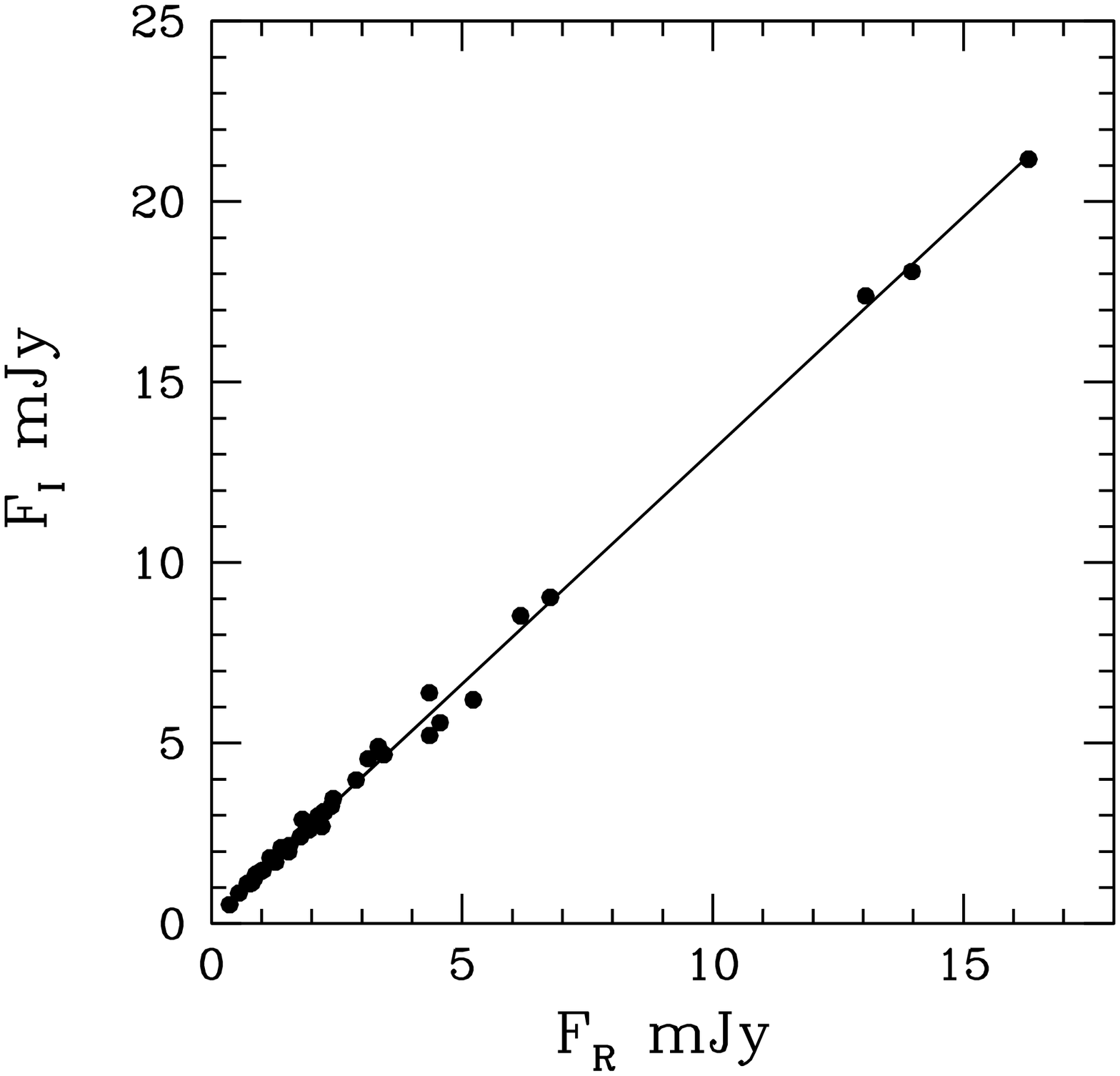}
\caption{Flux--flux dependence of the blazar S5 0954+658, where the  B (left), V (center)  and I (right)
band fluxes are plotted against the R flux; the lines represent linear regressions.}
 \end{figure*}

\subsection{BL Lacertae}

BL Lacertae is a prototype of the blazar class (at the redshift $z=0.069$; Miller \& Hawley 1977) and is classified as
low frequency peaked blazar (LBL).
BL Lac has long-known strong variability in optical bands and is one of the favorite targets of multi-wavelength
campaigns, such as those carried out by the Whole Earth Blazar Telescope (WEBT/GASP; B\"ottcher et al.\ 2003; Villata et al.\ 2009; Raiteri et al.\
2009, 2010, 2013 and references therein). It displays intense optical variability on short and intra-day
time-scales (e.g.\ Clements \& Carini 2001; Agarwal \& Gupta 2015; Gaur et al.\ 2014).  
It also shows strong polarization variability (Marscher et al.\ 2008; Gaur et al.\ 2014, and references therein).
In previous studies, it has been found that flux variations are  associated with  spectral changes (Villata et al.\ 2002; 
Papadakis et al.\ 2003; Hu et al.\ 2006).
Most of the observed properties of this source can be attributed to  synchrotron emission from highly relativistic electrons within
a helical jet (Raiteri et al.\ 2009).
BL Lac is within a relatively bright host galaxy and we removed its contribution from the
observed magnitudes by following the method provided in Gaur et al.\ (2015a).

\subsubsection{Long term flux and colour variability}

The LTV LC of BL Lac is shown in Figure 11 for a total of 73 nights over nearly
three years spanning  2014 through  2016. The source displayed significant flux changes on long-term time-scales
in all the bands we measured (B, V, R and I).  The variability amplitudes seen in the B, V, R and I bands are
 263, 260, 254, and 251 per cent, respectively. The equivalent total changes in magnitudes are $\Delta$B=1.31,
$\Delta$V=1.28, $\Delta$R=1.15 and $\Delta$I=1.12.

Our entire set of observations can be divided into three seasons. The first observation season covered the period 23 May
2014--30th September 2014, during which BL Lac was observed for a total of 14 nights.
 The second cycle spanned 20 February 2015--14 December 2015 when  the source was observed on 36 nights.
The third season covers the period from 30 June 2016--1 October 2016 when BL Lac was observed for a total of 23 nights.
The magnitude of BL Lac ranged between 13.4--12.3 in the R band during this three year span. In some of the nights, observations
were performed using two telescopes and we found good agreement between them. This source was extensively observed
in the past by the WEBT collaboration and they found the magnitude to vary between 14.8--12.5 in the R band (Villata et al.\ 
2002; 2004; Raiteri et al.\ 2010, 2013, and references therein). Hence, we can conclude that  we observed BL Lac in a relatively
high state, as noted above.

We calculated the spectral index for each night of our observations and they are shown in the lower panel of Fig.\ 12. The $\alpha_{BR}$
slope is very steep during the observing period and varies between 2.4--3.5. This plot again indicates that the emission is dominated by
synchrotron emission and shows strong bluer-when-brighter trends.

A mildly chromatic behaviour is found for the LTV, with a slope of $0.072\pm0.02$; intercept of $0.14\pm0.21$;
$r$=0.468; $p$=3.3e-05  for the (B$-$R) CI versus R magnitudes shown in Figure 13. This is in accordance with 
 Villata et al.\ (2004) where they found a similar slope
between (B-R) versus R during the period 1997--2002;  the mildly chromatic long-term variations were simply explained in terms of
changes in Doppler factors. Also, it can be seen from Fig.\ 12 that there are various outbursts  superimposed on the long
term trends which vary over time and the chromatism is also variable. The slope of 0.1  should be considered to
represent only a mean value.

It can seen that the the smaller flares/brightness changes on time-scales of days are superimposed on the longer term trends.
In  previous studies of BL Lacertae it was argued that much of the variability can be explained by changes       
in the Doppler factor of the jet emission region (e.g., Villata et al.\ 2004; Larionov et al.\ 2010). 

\begin{figure}
\centering
\includegraphics[width=8cm , angle=0]{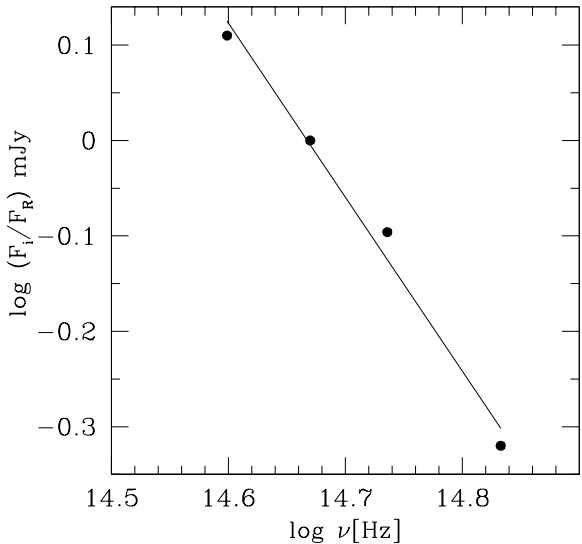}
\caption{Relative SED of  S5 0954+658, normalized to R band.}
 \end{figure}

\begin{figure*}
\centering
\includegraphics[width=12cm , angle=0]{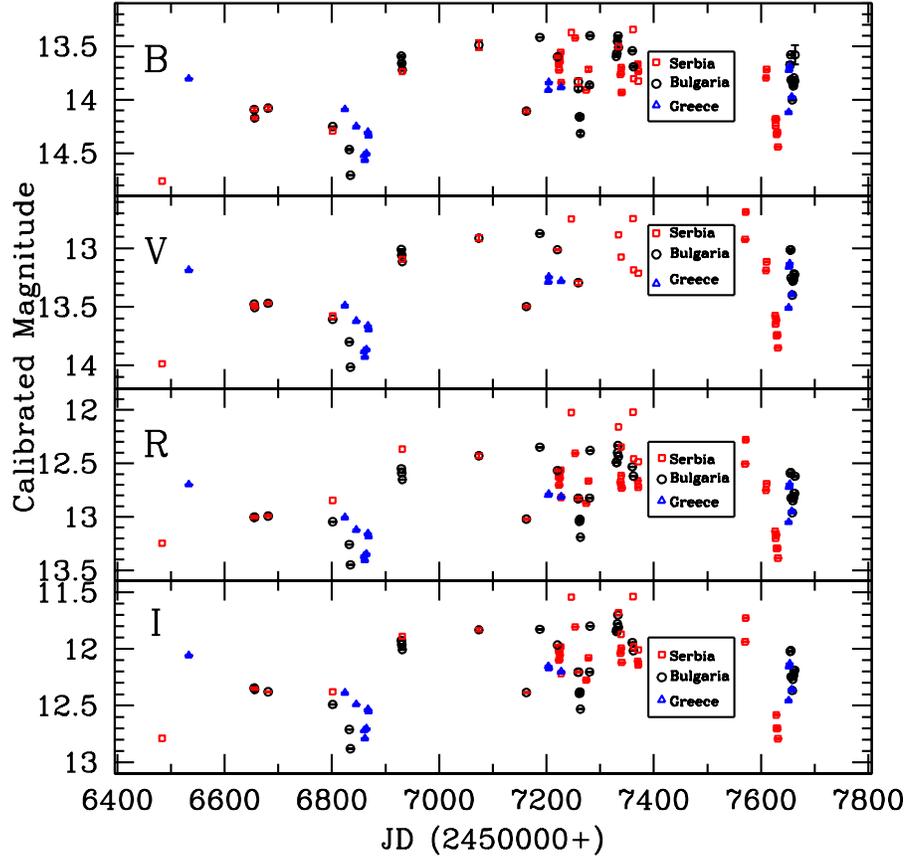}
\caption{Long term light curve of BL Lacertae during the period 2014--2016; labels as in Fig.\ 1.}
 \end{figure*}

\subsubsection{Spectral Energy Distribution}

We calculated the flux-flux dependencies for the B, V, R and I bands and these are shown in Figure 14. The relations are not
very well fit by linear functions and hence the spectra are far from  power-laws.  So we fit them with second order polynomials, 
i.e., $a+b \times F+c \times F^{2}$. We then produced continuum spectra of BL Lacetae by using these polynomial regressions and 
these are shown in Fig.\ 15 for R band fluxes of 15, 25 and 35 mJy, which are within the range of our measurements.

If the intrinsic spectrum is a power law, $F_{\nu} \propto \nu^{\alpha}$ and $\alpha$ is taken as independent of wavelength over
this modest range, then the flux-flux plots would have shown linear dependencies; however, this is not quite the case in our
observations as shown in Fig.\ 14.
Following Larionov et al.\ (2010), if we assume that the flux variations arise from changes in the viewing angle which lead to 
changes in the Doppler factor, $\delta$, then in the observer's rest frame, the flux density is $F_{\nu} = \delta^{p} F_{\nu}^{\prime}$ 
(Rybicki \& Lightman 1979; Urry \& Padovani 1995), where primed quantities refer to the source rest frame; $p=3$ for the 
situation where the radiation comes from a moving source with a compact emission zone (shock/knot), while $p=2$ is appropriate 
for a smooth continuous jet (Begelman et al.\ 1994). Also, the frequencies are affected by the doppler shift as $\nu$= 
$\delta$ $\nu^{\prime}$. Hence, in Fig. 15, a logarithmic plot, the variation in the Doppler factor by $\Delta$~log~$\delta$ 
leads to a shift in flux as well as in frequency in the following form: $\Delta$~log~$F_{\nu} = p~\Delta$~log~$\delta$ 
and $\Delta$~log~$\nu$ =$ \Delta$~log~$\delta$.
Here, we calculate the shift in the Doppler factor which is required to obtain the spectrum
corresponding to 15 mJy from the highest flux spectrum of 35 mJy. We can roughly reproduce (as shown by the coloured lines in Fig.\ 15)
the lower spectrum from the higher one by lowering $\delta$ by a factor of 1.19 for $p=3$ or by a factor of 1.25 assuming $p=2$,
but given the very modest curvature over our limited range of bands, we cannot really distinguish between these possibilities.
Hence, the most plausible explanation to describe the overall flux and colour variability during 2013--2017 
is in terms of modest Doppler factor
variations which affect the observed radiation coming from a jet. But, due to the limited range of frequencies,
we can not completely ruled out the possibility of other variability models.

Larionov et al.\ (2010) found a somewhat larger shift in the Doppler
factor by a factor of 1.58 with $p=3$ to fit a wider range of flux variations in data taken between 2000 and 2008.
Because they had measurements in more IR bands they argued that they could distinguish between the shock/knot and smooth jet
possibilities and favored the former.

\begin{figure*}
\centering
\includegraphics[width=9cm , angle=0]{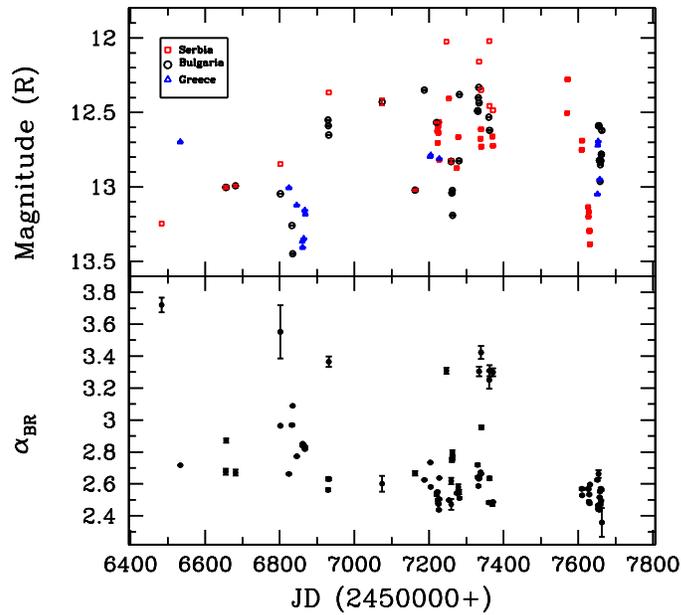}
\caption{ R band LC of BL Lac (upper panel) and the variation of spectral indices $\alpha_{BR}$ with respect to JD
 (lower panel).}
 \end{figure*}

\begin{figure}
\centering
\includegraphics[width=8cm , angle=0]{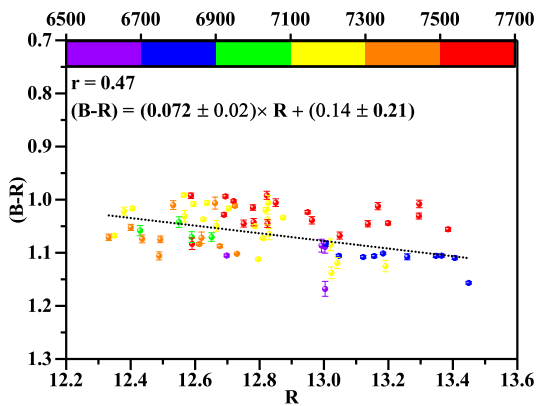}
\caption{(B-R) colour-index versus R magnitude of BL Lac on longer timescales}.
 \end{figure}

\section{Discussion and Conclusions}

We have presented the results of photometric monitoring of three TeV blazars, namely, 3C 66A, S4 0954+658 and BL Lacertae 
over nearly four years in 2013--2017 for a total of 536 nights in B, V, R and I bands. We studied their flux and spectral 
variability on these long timescales. All  three sources showed significant variability during our observing runs with 
the maximum variability shown by S5 0954+658 ($\Delta$R $\sim$4). BL Lac revealed a more moderate variation
of around $\Delta$R $\sim$1.12 while weaker variations of $\Delta$R $\sim$0.65 were displayed by 3C 66A. Comparison of our
photometric observations with the earlier observations of same sources indicates that we observed the blazars 3C 66A
and BL Lac in high states. However, the blazar S5 0954+658 was observed in low as well as high states. This source
underwent many strong flares during our observations and reached a maximum brightness of around $\sim$13 magnitude in February 2015.

We also studied the correlation between R magnitude and the colour indices.  We did not find any correlation
for 3C 66A but found weak positive correlations for S5 0954+658 and BL Lac. It can be seen that the spectra of S5 0954+658 
are getting flatter as the flux increases (Table 2) which indicates a bluer-when-brighter trend. For BL Lacertae, we found a 
positive slope of 0.1 between the (B$-$R) colour-index and R magnitude, again indicating a similar trend, which is 
commonly seen for LBL blazars. 
 Bluer-when-brighter chromatism can be explained with a one-component synchrotron model in that the more intense the energy
release, the higher the typical particle's energy and the higher the corresponding frequency (Fiorucci et al.\ 2004).
It can also be result from the injection of fresh electrons with an energy distribution harder than that of the previously
cooled ones (e.g., Kirk, Rieger \& Mastichiadis 1998; Mastichiadis \& Kirk 2002).
The long-term trends we saw are generally mildly chromatic. 
The flatter slope of the colour--magnitude diagram could be revealing the presence of multiple variable components, each evolving differently. 
Hence, superposition of different spectral slopes from multiple components leads to a general reduction of 
the colour--magnitude correlations (Bonning et al.\ 2012, Bachev 2015).
It has been found in  previous studies that the slope of the optical spectrum is often only weakly sensitive to the long-term trends
while strictly following the bluer when brighter trends on short timescales  (Villata et al.\ 2002; Larionov et al.\ 2010;
Gaur et al.\ 2012, 2015a). 

\begin{figure*}
\centering
\includegraphics[width=5cm , angle=0]{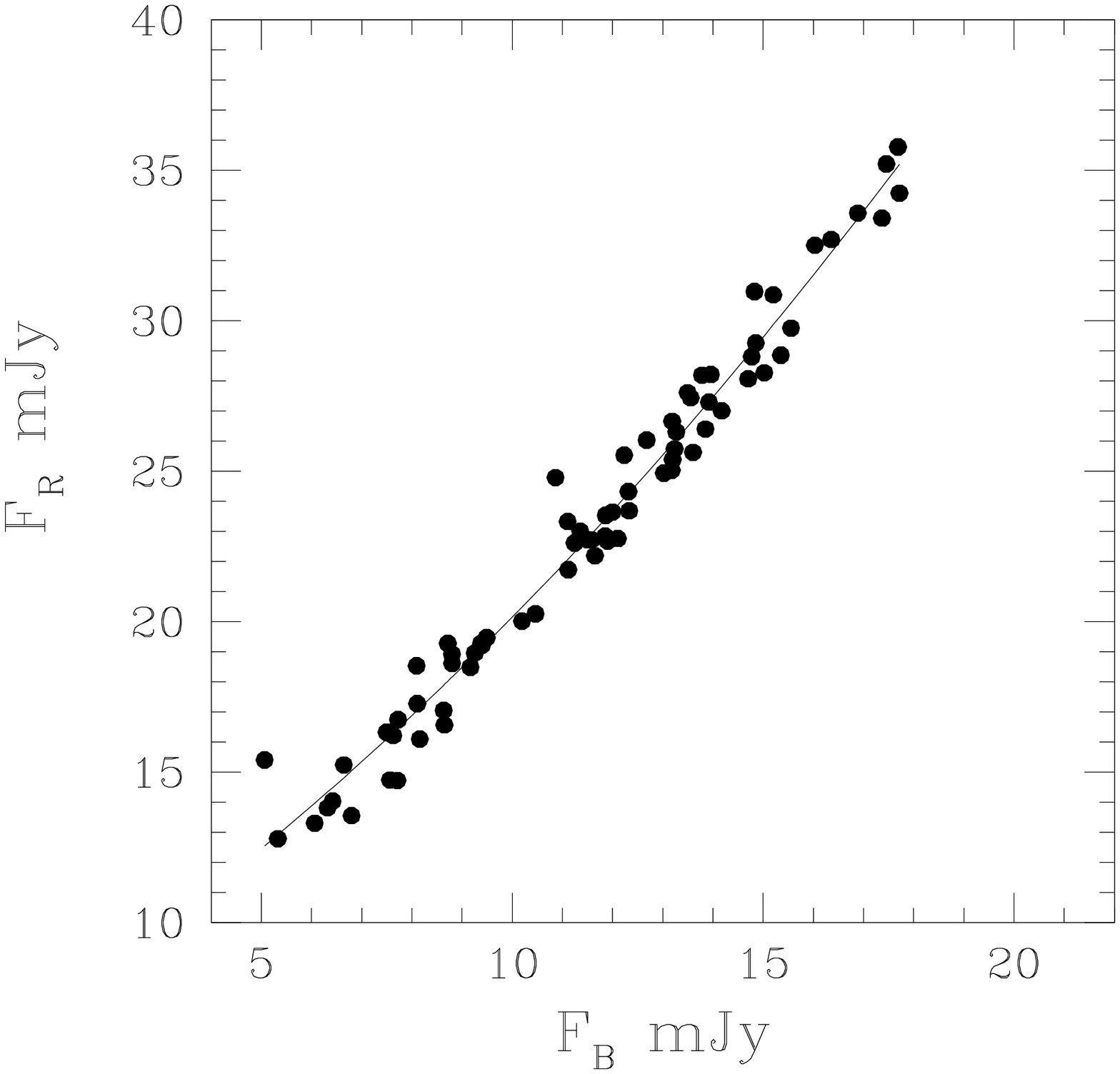}
\includegraphics[width=5cm , angle=0]{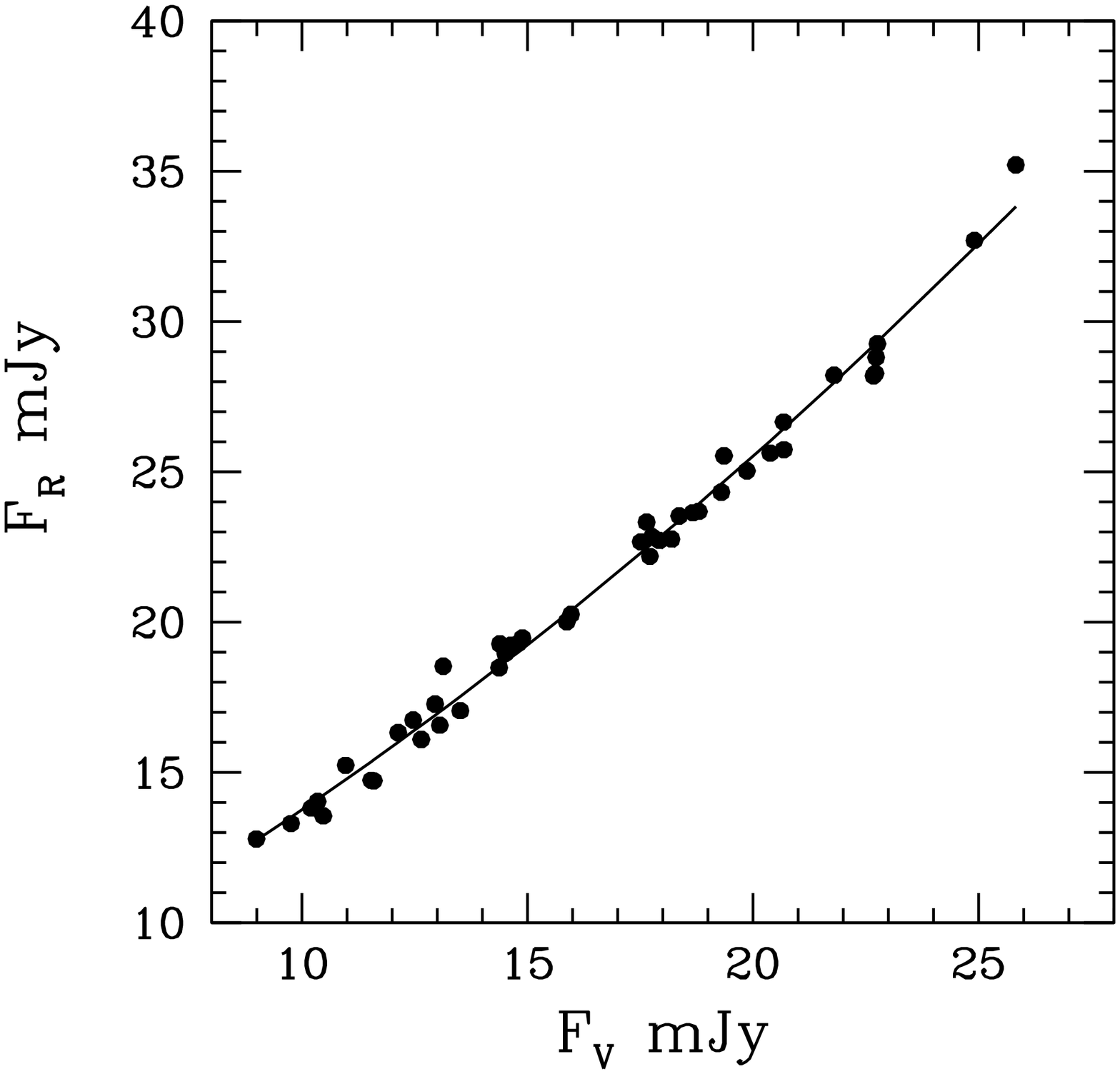}
\includegraphics[width=5cm , angle=0]{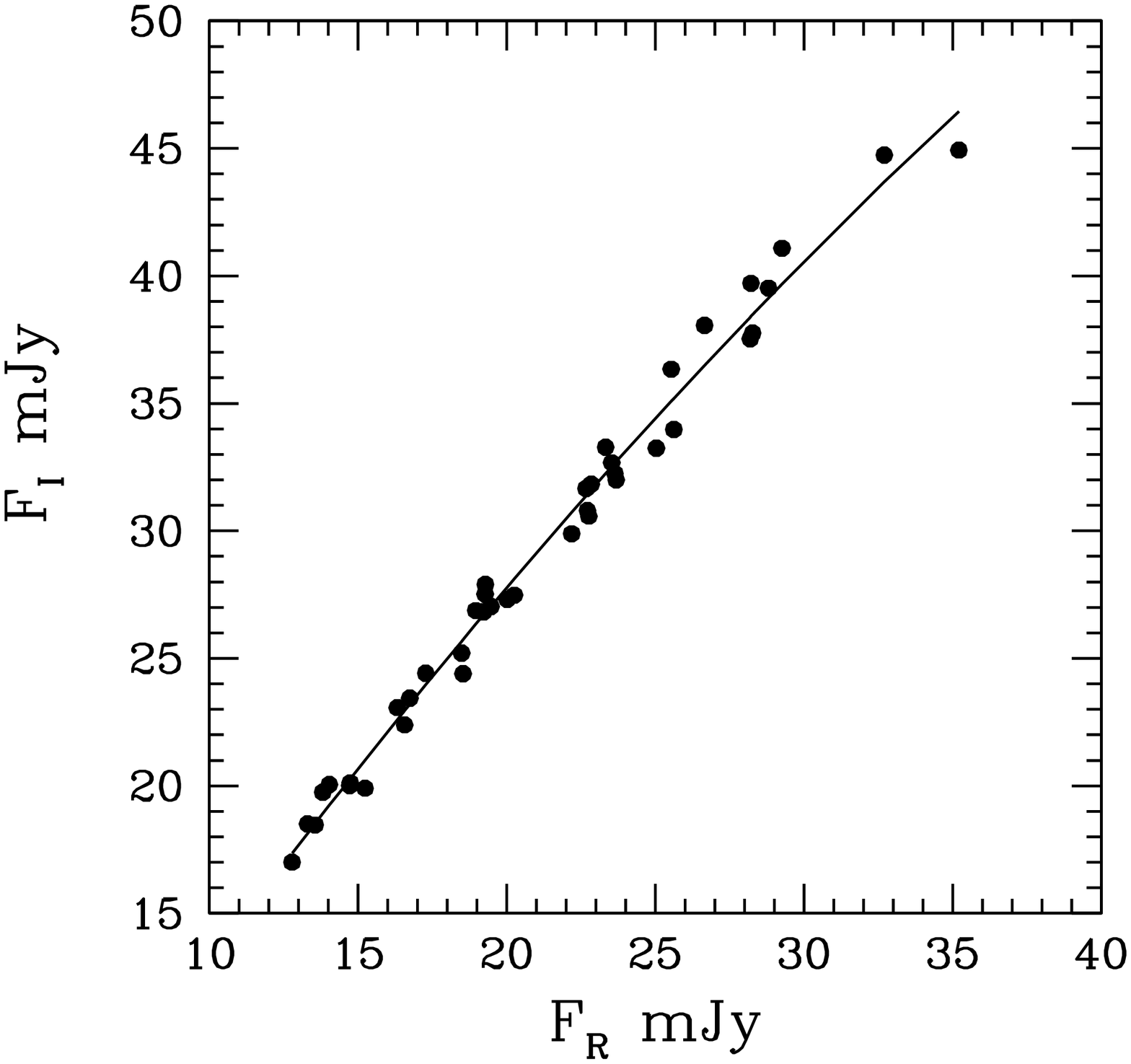}
\caption{Flux-flux dependence of the BL Lac, where the R flux is plotted against the B (left), V (center)  and I (right) band fluxes. The lines represent second-order polynomial regressions.}
 \end{figure*}

IDV of two of these sources, S5 0954+658 and BL Lac, were investigated in Bachev (2015) and Gaur et al.\  (2017), respectively.
S5 0954+658 was observed by Bachev (2015) during its unprecedented high state in 2015 February when it reached a maximum brightness
 of $\sim$13 magnitude. Rapid, violent variations of 0.1--0.2 mag h$^{-1}$ were observed on intra-day timescales and favour a
geometrical scenario to account for this variability.  The IDV of BL Lacertae was observed by Gaur et al.\ (2017) over a total of 45 nights 
and they found strong variability of $\sim$0.1 mag over a few hours on several nights. The IDV of BL Lacertae can be associated with models 
based on shocks moving through a turbulent plasma jet.

Models that have been proposed to explain optical variability of blazars generally involve one or more of the following processes: 
changes in the electron energy density distribution
of the relativistic particles producing variable synchrotron emission; inhomogeneities in magnetic fields; shocks 
accelerating particles in the bulk relativistic plasma jet; turbulence behind an outgoing shock in a jet; and
irregularities in the jet flows (e.g., Marscher, Gear
\& Travis 1992; Giannios et al.\ 2009; Marscher 2014; Gaur et al. 2014; Calafut \& Wiita 2015; Pollack et al.\ 2016).
The flux variations in blazars on longer timescales can also be explained by the launching of new components (or shocks) (e.g.,
Marscher \& Gear 1985, Hughes et al.\ 1989) or mechanisms involving geometrical effects
such as modestly swinging jets, for which the path of the relativistically moving source regions along the jet deviates a bit
from the line of sight, leading to a changing Doppler factor (e.g., Gopal-Krishna \& Wiita 1992; Pollack et al.\ 2016) or in a
helical jet model where the orientation changes though the rotation about the helix (Villata et al.\ 2009).

 In order to try to understand the origin of variability of these blazars during our observations, 
we studied flux-flux relations of these sources, based on the assumption that linear relations between observed fluxes
at different wavelengths suggest that the slope does not change during the period of flux variability. And, if the variability
arises from a single variable source (or multiple components with similar SEDs) then its slope can be derived from the 
slopes of the flux--flux relations. Hence, we calculated the flux--flux relations of these three blazars and found that 
those of 3C 66A and S5 0954+658  followed linear regressions. Hence, a single variable source (or
multiple components with similar SEDs) are probably responsible for the variability we are seeing. Using the slopes of the
flux--flux relations, we constructed the relative SED of this variable components. The slopes of the 
variable emitting region of 3C 66A and S5 0954+658 are found to be $\simeq1.65$ and $1.8$, respectively. Hence, using the flux--flux 
relations we suggest that the variability in these two sources can be interpreted in terms of a non-thermal
variable component which has a continuous injection of relativistic electrons with a power law energy distributions around
4.3 and 4.6, respectively. 

The flux--flux relations of BL Lac are better fitted with a second order
polynomial which implies that the observed flux variations more likely arise from the changes in the Doppler factor of  
the relativistic jet, as suggested earlier  (Villata et al.\ 2002; Larionov et al.\ 2010, and references therein).
Hence, we calculated the optical spectra of BL Lac at different flux states between 15 mJy and 35 mJy. 
We found that the lowest spectrum could be reasonably reproduced from the highest flux spectrum by decreasing the Doppler 
factor by a factor of $\sim$1.2. These observations of long-term flux and mildly 
achromatic BL Lac variability might be explained to arise from  changes in the viewing angle, which lead to the Doppler factor
becoming variable (Larionov et al.\ 2010 and references therein).

Our observations of these three blazars allow us to make plausible  inferences about the origins of variability in optical bands.
However, as  we are dealing
with very narrow range of frequencies in the optical SEDs and we have limited data set which is simultaneous in B, V, R and I bands needed to properly study
flux--flux relations the conclusions must be used with caution. Even larger datasets and broader multi-wavelength modelling of SEDs
 are required to firmly constrain various blazar jet parameters.

\begin{figure}
\centering
\includegraphics[width=8cm , angle=0]{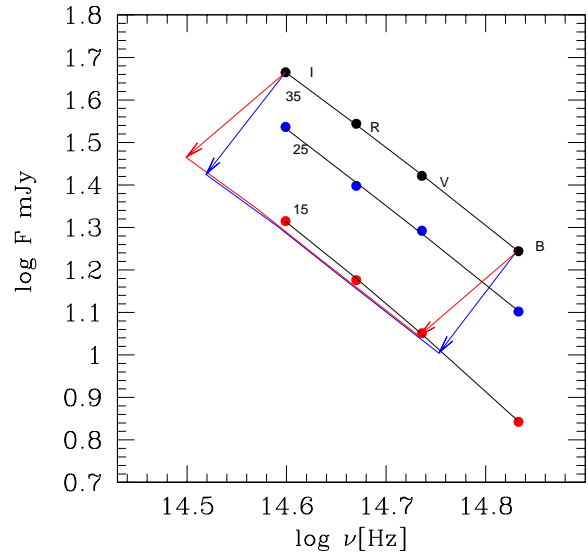}
\caption{Spectra of  BL Lac in optical bands obtained from polynomial regressions. Black, blue and red curves connecting the I, R, V and B bands represent the spectra for 35, 25 and 15 mJy, respectively. The blue and red arrows and the corresponding curves connecting them represents the spectra obtained from the 35 mJy spectrum corresponding
to decreasing $\delta$ by a factor of 1.19 for $p=3$ (blue) or by a factor of $1.25$ for $p=2$ (red). Details are  in the text.  }
 \end{figure}

 We would like to thanks the anonymous reviewer for the constructive comments which helped us to improve the manuscript 
scientifically.
HG acknowledges the financial support from the Department of Science \& Technology, INDIA
through INSPIRE faculty award IFA17-PH197  at ARIES, Nainital.
This research was partially supported by the Bulgarian National
Science Fund of the Ministry of Education and Science
under grants DN 08-1/2016, DM 08-2/2016, DN
18-13/2017, DN 18-10/2017 and KP-06-H28/3 (2018)
as well as by the project RD-08-112/2018 of the
University of Shumen.
The authors thank the 
Director of Skinakas Observatory Prof.\ I.\ Papamastorakis and Prof.\ I.\ Papadakis for the award of telescope time.
The Skinakas Observatory is a collaborative project of the University of Crete, the Foundation for Research and Technology -- 
Hellas, and the Max-Planck-Institut f\"ur Extraterrestrische Physik. 
The Abastumani team acknowledges financial support by the Shota Rustaveli National Science Foundation of Georgia
under contract FR/217554/16.
GD and OV gratefully acknowledge the observing grant support from the Institute of Astronomy and Rozhen
National Astronomical Observatory, Bulgarian Academy of Sciences, via bilateral joint research project ``Observations of ICRF
radio-sources visible in optical domain''. This work is a part of the Projects No.\
176011 (``Dynamics and kinematics of celestial bodies and systems"), No. 176004 (``Stellar physics") and No.\ 176021 (``Visible
and invisible matter in nearby galaxies: theory and observations") supported by the Ministry of Education,
Science and Technological Development of the Republic of Serbia.

{}

\newpage`
\Huge
\vspace*{4.5in}
\begin{center}
{\bf Supplement Material}
\end{center}
 
\clearpage

\begin{table*}
{\bf Table 1.} Observation log of optical photometric observations of the blazar 3C 66A.
\centering
\noindent
\scalebox{.90}{
\setlength{\tabcolsep}{0.020in}
\begin{tabular}{cccccccccccc} \hline \hline
Date of     &Telescope  &Number  &Date of   &Telescope  &Number  &Date of  &Telescope  &Number &Date of   &Telescope  &Number\\
Observation &           &B,V,R,I  &Observation     &           &B,V,R,I        &Observation &        &B,V,R,I   &Observation  & &B,V,R,I      \\ \hline
11.01.2013    &H        &0,0,1,0          &12.01.2014  &F       &0,0,3,0        &22.12.2014  &H &0,0,1,0        &11.01.2016  &H &0,0,1,0\\
12.01.2013    &H        &0,0,1,0          &21.01.2014  &F       &0,0,3,0        &23.12.2014  &H &0,0,1,0        &12.01.2016  &D &2,0,2,2\\
15.01.2013    &F        &0,0,4,0          &03.02.2014  &F       &0,0,6,0        &25.12.2014  &F &0,0,3,0        &12.01.2016  &H &0,0,1,0\\
15.01.2013    &H        &0,0,1,0          &03.02.2014  &H       &0,0,1,0        &26.12.2014  &H &0,0,1,0        &13.01.2016  &H &0,0,1,0\\
16.01.2013    &H        &0,0,1,0          &04.02.2014  &H       &0,0,1,0        &27.12.2014  &H &0,0,1,0        &14.01.2016  &H &0,0,1,0\\
17.01.2013    &H        &0,0,1,0          &05.02.2014  &H       &0,0,1,0        &08.12.2015  &F &0,0,4,0        &17.01.2016  &H &0,0,1,0\\
18.01.2013    &H        &0,0,1,0          &07.02.2014  &H       &0,0,1,0        &19.01.2015  &F &0,0,4,0        &20.01.2016  &F &0,0,4,0\\
19.01.2013    &H        &0,0,1,0          &08.02.2014  &F       &0,0,4,0        &19.01.2015  &H &0,0,1,0        &04.02.2016  &F &0,0,5,0\\
24.01.2013    &F        &0,0,6,0          &08.02.2014  &H       &0,0,1,0        &21.01.2015  &H &0,0,1,0        &06.02.2016  &D &9,0,9,9\\
01.02.2013    &H        &0,0,1,0          &19.02.2014  &F       &0,0,5,0        &22.01.2015  &H &0,0,1,0        &07.02.2016  &D &3,0,3,3\\
04.02.2013    &F        &0,0,6,0          &23.02.2014  &F       &0,0,3,0        &25.01.2015  &F &0,0,4,0        &08.02.2016  &H &0,0,1,0\\
04.02.2013    &H        &0,0,1,0          &24.02.2014  &H       &0,0,1,0        &02.02.2015  &F &0,0,5,0        &09.02.2016  &H &0,0,1,0\\
05.02.2013    &H        &0,0,1,0          &26.02.2014  &H       &0,0,1,0        &11.02.0215  &H &0,0,1,0        &11.02.2016  &H &0,0,1,0\\
06.02.2013    &H        &0,0,1,0          &27.02.2014  &H       &0,0,1,0        &12.02.2015  &H &0,0,1,0        &12.02.2016  &H &0,0,1,0\\
07.02.2013    &F        &0,0,5,0          &28.02.2014  &H       &0,0,1,0        &16.02.2015  &H &0,0,1,0        &15.02.2016  &F &0,0,5,0\\
07.02.2013    &H        &0,0,1,0          &09.03.2014  &F       &0,0,4,0        &21.02.2015  &F &0,0,4,0        &15.02.2016  &H &0,0,1,0\\
08.02.2013    &H        &0,0,1,0          &19.03.2014  &F       &0,0,5,0        &03.03.2015  &F &0,0,4,0        &19.02.2016  &F &0,0,4,0\\
04.03.2013    &H        &0,0,1,0          &27.03.2015  &H       &0,0,1,0        &08.03.2015  &F &0,0,5,0        &01.03.2016  &F &0,0,4,0\\
06.03.2013    &H        &0,0,1,0          &28.03.2015  &H       &0,0,1,0        &16.03.2015  &H &0,0,1,0        &04.03.2016  &F &0,0,5,0\\
07.03.2013    &F        &0,0,6,0          &29.03.2015  &H       &0,0,1,0        &19.03.2015  &F &0,0,4,0        &10.03.2016  &H &0,0,1,0\\
10.03.2013    &H        &0,0,1,0          &30.03.2015  &H       &0,0,1,0        &23.03.2015  &H &0,0,1,0        &12.03.2016  &H &0,0,1,0\\
11.03.2013    &H        &0,0,1,0          &31.03.2015  &H       &0,0,1,0        &31.03.2015  &F &0,0,5,0        &26.06.2016  &F &0,0,3,0\\
12.03.2013    &H        &0,0,1,0          &02.04.2014  &F       &0,0,2,0        &16.04.2015  &F &0,0,4,0        &18.07.2016  &F &0,0,4,0\\
21.03.2013    &F        &0,0,6,0          &13.06.2014  &F       &0,0,4,0        &16.06.2015  &H &0,0,1,0        &22.07.2016  &F &0,0,5,0\\
04.04.2013    &F        &0,0,8,0          &15.06.2014  &B       &3,3,3,3        &18.06.2015  &H &0,0,1,0        &23.07.2016  &H &0,0,1,0\\
28.06.2013    &F        &0,0,6,0          &24.06.2014  &H       &0,0,1,0        &29.06.2015  &B &3,3,3,3        &25.07.2016  &H &0,0,1,0\\
12.97.2013    &F        &0,0,4,0          &27.06.2014  &F       &0,0,4,0        &30.06.2015  &B &3,3,3,3        &01.08.2016  &F &0,0,3,0\\
26.07.2013    &B        &3,3,3,3          &06.07.2014  &B       &3,3,3,3        &08.07.2015  &F &0,0,1,0        &18.08.2016  &F &0,0,4,0\\
26.07.2013    &F        &0,0,6,0          &08.07.2014  &F       &0,0,3,0        &11.07.2015  &F &0,0,4,0        &29.08.2016  &F &0,0,3,0\\
06.08.2013    &F        &0,0,4,0          &21.07.2014  &B       &3,3,3,3        &15.07.2015  &H &0,0,1,0        &29.08.2016  &H &0,0,1,0\\
11.08.2013    &F        &0,0,4,0          &21.07.2014  &F       &0,0,4,0        &20.07.2015  &H &0,0,1,0        &30.08.2016  &H &0,0,1,0\\
19.08.2013    &F        &0,0,5,0          &22.07.2014  &B       &3,3,3,3        &21.07.2015  &H &0,0,1,0        &02.09.2016  &H &0,0,1,0\\
28.08.2013    &B        &3,3,3,3          &24.07.2014  &H       &0,0,1,0        &22.07.2015  &F &0,0,5,0        &03.09.2016  &H &0,0,1,0\\
30.08.2013    &F        &0,0,5,0          &28.07.2014  &B       &3,3,3,3        &23.07.2015  &B &5,5,5,5        &04.09.2016  &H &0,0,1,0\\
04.09.2013    &H        &0,0,1,0          &29.07.2014  &B       &3,3,3,3        &03.08.2015  &F &0,0,3,0        &05.09.2016  &H &0,0,1,0\\
05.09.2013    &H        &0,0,1,0          &07.08.2014  &F       &0,0,5,0        &17.08.2015  &F &0,0,3,0        &12.09.2016  &F &0,0,4,0\\
06.09.2013    &E        &1,1,1,1          &14.08.2014  &F       &0,0,4,0        &31.08.2015  &F &0,0,4,0        &18.09.2016  &F &0,0,4,0\\
06.09.2013    &F        &0,0,8,0          &21.08.2014  &F       &0,0,3,0        &14.09.2015  &F &0,0,4,0        &20.09.2016  &B &5,5,5,5\\
10.09.2013    &H        &0,0,1,0          &01.09.2014  &F       &0,0,4,0        &17.09.2015  &H &0,0,1,0        &21.09.2016  &B &5,5,5,5\\
11.09.2013    &H        &0,0,1,0          &07.09.2014  &F       &0,0,5,0        &19.09.2015  &H &0,0,1,0        &24.09.2016  &H &0,0,1,0\\
18.09.2013    &F        &0,0,6,0          &16.09.2014  &F       &0,0,4,0        &20.09.2015  &F &0,0,4,0        &26.09.2016  &F &0,0,4,0\\
30.09.2013    &F        &0,0,5,0          &19.09.2014  &H       &0,0,1,0        &28.09.2015  &F &0,0,5,0        &01.10.2016  &F &0,0,4,0\\
04.10.2013    &H        &0,0,1,0          &20.09.2014  &H       &0,0,1,0        &04.10.2015  &F &0,0,5,0        &10.10.2016  &F &0,0,4,0\\
05.10.2013    &F        &0,0,5,0          &21.09.2014  &F       &0,0,4,0        &10.10.2015  &H &0,0,1,0        &24.10.2016  &H &0,0,1,0\\
05.10.2013    &H        &0,0,1,0          &21.09.2014  &H       &0,0,1,0        &12.10.2015  &H &0,0,1,0        &25.10.2016  &H &0,0,1,0\\
06.10.2013    &H        &0,0,1,0          &22.09.2014  &H       &0,0,1,0        &13.10.2015  &H &0,0,1,0        &26.10.2016  &H &0,0,1,0\\
11.10.2013    &H        &0,0,1,0          &23.09.2014  &H       &0,0,1,0        &14.10.2015  &H &0,0,1,0        &27.10.2016  &F &0,0,3,0\\
12.10.2013    &H        &0,0,1,0          &24.09.2014  &H       &0,0,1,0        &15.10.2015  &H &0,0,1,0        &28.10.2016  &H &0,0,1,0\\
14.10.2013    &F        &0,0,7,0          &25.09.2014  &H       &0,0,1,0        &29.10.2015  &F &0,0,4,0        &29.10.2016  &H &0,0,1,0\\
28.10.2013    &F        &0,0,7,0          &26.09.2014  &H       &0,0,1,0        &09.11.2015  &F &0,0,5,0        &07.11.2016  &F &0,0,4,0\\
02.11.2013    &F        &0,0,4,0          &01.10.2014  &F       &0,0,4,0        &09.11.2015  &H &0,0,1,0        &21.11.2016  &F &0,0,4,0\\
14.11.2013    &F        &0,0,6,0          &13.10.2014  &F       &0,0,3,0        &10.11.2015  &H &0,0,1,0        &24.11.2016  &H &0,0,1,0\\
24.11.2013    &H        &0,0,1,0          &19.10.2014  &F       &0,0,4,0        &11.11.2015  &H &0,0,1,0        &27.11.2016  &F &0,0,2,0\\
25.11.2013    &F        &0,0,5,0          &21.10.2014  &H       &0,0,1,0        &12.11.2015  &H &0,0,1,0        &29.11.2016  &H &0,0,1,0\\
25.11.2013    &H        &0,0,1,0          &22.10.2014  &H       &0,0,1,0        &13.11.2015  &D &3,0,3,2        &30.11.2016  &H &0,0,1,0\\
27.11.2013    &H        &0,0,1,0          &23.10.2014  &H       &0,0,1,0        &13.11.2015  &H &0,0,1,0        &05.12.2016  &F &0,0,4,0\\
28.11.2013    &H        &0,0,1,0          &24.10.2014  &H       &0,0,1,0        &14.11.2015  &F &0,0,5,0        &20.12.2016  &F &0,0,5,0\\
30.11.2013    &F        &0,0,6,0          &25.10.2014  &H       &0,0,1,0        &17.11.2015  &H &0,0,1,0        &24.12.2016  &F &0,0,2,0\\
13.12.2013    &F        &0,0,4,0          &30.10.2014  &F       &0,0,4,0        &23.11.2015  &F &0,0,4,0        &26.12.2016  &H &0,0,1,0\\
23.12.2013    &H        &0,0,1,0          &10.11.2014  &F       &0,0,3,0        &28.11.2015  &F &0,0,4,0        &02.01.2016  &H &0,0,1,0\\
24.12.2013    &F        &0,0,4,0          &16.11.2014  &F       &0,0,5,0        &07.12.2015  &F &0,0,4,0        &11.01.2017  &H &0,0,1,0\\
24.12.2013    &H        &0,0,1,0          &21.11.2014  &H       &0,0,1,0        &07.12.2015  &H &0,0,1,0        &26.01.2017  &H &0,0,1,0\\
25.12.2013    &H        &0,0,1,0          &22.11.2014  &H       &0,0,1,0        &08.12.2015  &H &0,0,1,0        &27.01.2017  &H &0,0,1,0\\
26.12.2013    &H        &0,0,1,0          &23.11.2014  &H       &0,0,1,0        &12.12.2015  &D &2,0,2,2        &28.01.2017  &H &0,0,1,0\\
27.12.2013    &H        &0,0,1,0          &24.11.2014  &H       &0,0,1,0        &13.12.2015  &D &2,0,2,2        &29.01.2017  &H &0,0,1,0\\
29.12.2013    &F        &0,0,4,0          &25.11.2014  &F       &0,0,4,0        &13.12.2015  &H &0,0,1,0        &30.01.2017  &H &0,0,1,0\\
29.12.2013    &H        &0,0,1,0          &25.11.2014  &H       &0,0,1,0        &15.12.2015  &H &0,0,1,0        &31.01.2017  &H &0,0,1,0\\
30.12.2013    &H        &0,0,1,0          &26.11.2014  &H       &0,0,1,0        &18.12.2015  &F &0,0,5,0        &24.02.2017  &H &0,0,1,0\\
31.12.2013    &H        &0,0,1,0          &27.11.2014  &H       &0,0,1,0        &25.12.2015  &F &0,0,5,0        &26.02.2017  &H &0,0,1,0\\
01.01.2014    &H        &0,0,1,0          &28.11.2014  &H       &0,0,1,0        &06.01.2016  &F &0,0,4,0        &01.03.2017  &H &0,0,1,0\\ \hline

\end{tabular}} \\
\end{table*}

\begin{table*}
\centering
\noindent
\setlength{\tabcolsep}{0.010in}
\begin{tabular}{cccccccccccc} \hline \hline
Date of     &Telescope  &Number  &Date of   &Telescope  &Number  &Date of  &Telescope   &Number &Date of &Telescope   &Number\\
Observation &           &B,V,R,I  &Observation     &           &B,V,R,I        &Observation &   &B,V,R,I  &Observation &   &B,V,R,I \\ \hline

08.03.2017    &H        &0,0,1,0  &09.03.2017    &H   &0,0,1,0   &10.03.2017    &H  &0,0,1,0  &11.03.2017    &H        &0,0,1,0  \\
12.03.2017    &H        &0,0,1,0   &13.03.2017    &H  &0,0,1,0  &28.03.2017    &H        &0,0,1,0   &29.03.2017    &H  &0,0,1,0   \\
14.06.2017    &H        &0,0,1,0   &28.06.2017    &H  &0,0,1,0   &29.06.2017    &H        &0,0,1,0   &30.06.2017    &H  &0,0,1,0   \\
01.07.2017    &H        &0,0,1,0   \\ \hline

\end{tabular} \\
\end{table*}

\begin{table*}
 Observation log of optical photometric observations of the blazar S5 0954$+$658.
\centering
\noindent
\scalebox{.90}{
\setlength{\tabcolsep}{0.020in}
\begin{tabular}{cccccccccccc} \hline \hline
Date of     &Telescope  &Number  &Date of   &Telescope  &Number  &Date of  &Telescope   &Number &Date of &Telescope   &Number\\
Observation &           &B,V,R,I  &Observation     &           &B,V,R,I        &Observation &   &B,V,R,I  &Observation &   &B,V,R,I \\ \hline
05.02.2013  &F    &0,0,4,0    &11.05.2015  &F    &0,0,4,0      &19.02.2016 &G    &0,1,0,0    &24.08.2016 &G   &0,1,0,0               \\
01.05.2013  &F    &0,0,3,0    &13.05.2015  &E    &1,1,1,1      &19.02.2016 &F    &0,0,8,0    &26.08.2016 &G   &0,1,0,0               \\
09.05.2013  &F    &0,0,4,0    &18.05.2015  &A    &1,1,3,1      &20.02.2016 &G    &0,1,0,0    &08.10.2016 &G   &0,1,0,0               \\
17.10.2013  &F    &0,0,4,0    &27.05.2015  &F    &0,0,4,0      &29.02.2016 &F    &0,0,4,0    &10.11.2016 &F   &0,0,4,0               \\
02.11.2013  &F    &0,0,4,0    &12.06.2015  &F    &0,0,4,0      &01.03.2016 &D    &2,2,3,2    &23.11.2016 &F   &0,0,4,0               \\
03.12.2013  &E    &1,1,1,1    &12.06.2016  &D    &3,0,2,2      &02.03.2016 &D    &2,30,30,28 &02.12.2016 &G   &0,1,0,0               \\
13.12.2013  &F    &0,0,5,0    &13.06.2016  &D    &3,0,2,2      &06.03.2016 &E    &1,1,1,1    &06.12.2016 &F   &0,0,4,0               \\
25.12.2013  &F    &0,0,4,0    &14.06.2015  &D    &23,0,22,24   &12.03.2016 &G    &0,1,0,0    &10.12.2016 &G   &0,1,0,0               \\
08.01.2014  &F    &0,0,4,0    &15.06.2015  &D    &1,0,2,1      &18.03.2016 &F    &0,0,5,0    &21.12.2016 &F   &0,0,4,0               \\
04.02.2014  &F    &0,0,4,0    &16.06.2015  &D    &2,0,2,2      &24.03.2016 &G    &0,1,0,0    &28.12.2016 &G   &0,1,0,0               \\
20.02.2014  &F    &0,0,4,0    &19.06.2015  &D    &2,0,2,2      &30.03.2016 &F    &0,0,3,0    &30.12.2016 &G   &0,1,0,0               \\
09.03.2014  &F    &0,0,4,0    &26.06.2015  &E    &1,1,1,1      &01.04.2016 &G    &0,1,0,0    &06.01.2017 &G   &0,1,0,0               \\
21.03.2014  &F    &0,0,4,0    &19.07.2015  &D    &2,0,2,2      &06.04.2016 &D    &2,25.24,25 &07.01.2017 &G   &0,1,0,0               \\
01.04.2014  &F    &0,0,4,0    &24.07.2015  &F    &0,0,4,0      &07.04.2016 &D    &2,30,30,24 &13.01.2017 &G   &0,1,0,0               \\
16.04.2014  &F    &0,0,3,0    &17.10.2015  &F    &0,0,4,0      &11.04.2016 &F    &0,0,4,0    &29.01.2017 &C   &2,2,6,2               \\
30.04.2014  &F    &0,0,3,0    &29.10.2015  &F    &0,0,4,0      &12.04.2016 &D    &10,51,49,50&01.02.2017 &C   &2,2,4,2               \\
14.10.2014  &F    &0,0,2,0    &30.10.2015  &G    &0,1,0,0      &15.01.2016 &F    &0,0,5,0    &02.02.2017 &C   &2,2,4,2               \\
25.11.2014  &F    &0,0,4,0    &31.10.2015  &G    &0,1,0,0      &15.04.2016 &D    &8,40,42,42 &15.02.2017 &A   &1,1,2,1               \\
22.12.2014  &F    &0,0,5,0    &06.11.2015  &G    &0,1,0,0      &16.04.2016 &D    &7,53,52,52 &17.02.2017 &G   &0,1,0,0               \\
28.12.2014  &F    &0,0,6,0    &08.11.2015  &G    &0,1,0,0      &30.04.2016 &G    &0,1,0,0    &18.02.2017 &G   &0,1,0,0               \\
19.01.2015  &F    &0,0,4,0    &04.12.2015  &G    &0,1,0,0      &01.05.2016 &G    &0,1,0,0    &24.02.2017 &G   &0,1,0,0               \\
02.02.2015  &F    &0,0,5,0    &05.12.2015  &G    &0,1,0,0      &06.05.2016 &D    &1,3,2,3    &25.02.2017 &G   &0,1,0,0               \\
03.02.2015  &F    &0,0,6,0    &05.12.2015  &E    &1,1,1,1      &09.05.2016 &F    &0,0,4,0    &10.03.2017 &G   &0,1,0,0               \\
04.02.2015  &F    &0,0,5,0    &07.12.2015  &F    &0,0,6,0      &20.05.2016 &G    &0,1,0,0    &11.03.2017 &G   &0,1,0,0               \\
11.02.2015  &D    &20,0,19,20 &08.12.2015  &D    &4,0,2,2      &25.05.2016 &F    &0,0,4,0    &17.03.2017 &G   &0,1,0,0               \\
12.02.2015  &D    &40,0,40,40 &11.12.2015  &G    &0,2,0,0      &27.05.2016 &G    &0,1,0,0    &18.03.2017 &G   &0,2,0,0               \\
13.02.2015  &D    &50,0,50,50 &12.12.2015  &G    &0,1,0,0      &28.05.2016 &G    &0,1,0,0    &19.03.2017 &G   &0,2,0,0               \\
14.02.2015  &D    &36,0,37,37 &12.12.2015  &F    &0,0,8,0      &11.06.2016 &G    &0,1,0,0    &24.03.2017 &G   &0,1,0,0               \\
16.02.2015  &F    &0,0,5,0    &13.12.2015  &D    &29,0,30,30   &24.06.2016 &G   &0,1,0,0     &31.03.2017 &G   &0,1,0,0               \\
18.02.2015  &E    &1,1,1,1    &13.12.2015  &G    &0,1,0,0      &25.06.2016 &G   &0,1,0,0     &01.04.2017 &G   &0,1,0,0               \\
20.02.2015  &A    &0,0,3,0    &14.12.2015  &D    &0,0,1,0      &01.07.2016 &G   &0,1,0,0     &07.04.2017 &G   &0,1,0,0               \\
02.03.2015  &F    &0,0,6,0    &13.12.2015  &E    &1,1,1,1      &02.07.2016 &G   &0,1,0,0     &08.04.2017 &G   &0,1,0,0               \\
08.03.2015  &F    &0,0,6,0    &21.12.2015  &F    &0,0,6,0      &08.07.2016 &G   &0,1,0,0     &21.04.2017 &G   &0,1,0,0               \\
18.03.2015  &F    &0,0,5,0    &22.12.2015  &F    &0,0,6,0      &15.07.2016 &G   &0,1,0,0     &22.04.2017 &G   &0,1,0,0               \\
19.03.2015  &F    &0,0,4,0    &12.01.2016  &D    &20,0,19,20   &24.07.2016 &G   &0,1,0,0     &28.04.2017 &G   &0,1,0,0               \\
31.03.2015  &F    &0,0,4,0    &15.01.2016  &G    &0,1,0,0      &30.07.2016 &G   &0,1,0,0     &13.05.2017 &G   &0,1,0,0               \\
14.04.2015  &F    &0,0,4,0    &16.01.2016  &G    &0,1,0,0      &31.07.2016 &G   &0,1,0,0     &14.05.2017 &G   &0,1,0,0               \\
16.04.2015  &E    &1,1,1,1    &20.01.2016  &F    &0,0,4,0      &06.08.2016 &G   &0,1,0,0     &26.05.2017 &G   &0,1,0,0               \\
18.04.2015  &D    &1,1,1,1    &30.01.2016  &G    &0,1,0,0      &07.08.2016 &G   &0,1,0,0     &09.06.2017 &G   &0,1,0,0               \\
19.04.2015  &D    &2,0,2,2    &03.02.2016  &F    &0,0,3,0      &18.08.2016 &G   &0,1,0,0     &10.06.2017 &G   &0,1,0,0               \\
22.04.2015  &D    &3,0,2,2    &05.02.2016  &G    &0,1,0,0      &20.08.2016 &G   &0,1,0,0     &16.06.2017 &G   &0,1,0,0               \\
23.04.2015  &A    &1,1,3,1    &06.02.2016  &E    &1,1,1,1      &21.08.2016 &G   &0,1,0,0     &   &  & \\
25.04.2015  &D    &2,0,2,2    &07.02.2016  &E    &1,1,1,1      &22.08.2016 &G   &0,1,0,0     &   &  & \\\hline
\end{tabular}} 
\end{table*}

\begin{table*}
 Observation log of optical photometric observations of the blazar BL Lacertae.
\noindent
\setlength{\tabcolsep}{0.010in}
\begin{tabular}{cccccccccccc} \hline \hline
Date of        &Telescope   &Number & Date of        &Telescope   & Number & Date of        &Telescope   &Number & Date of        &Telescope   & Number \\
Observation    &            &B,V,R,I& Observation    &            &B,V,R,I & Observation    &            &B,V,R,I& Observation    &            &B,V,R,I \\ \hline
23.05.2014  &E  &1,1,1,1&       16.07.2015 &A  &2,2,2,2         & 04.11.2015 &D  &26,0,28,28    & 25.08.2016 &D  &13,14,28,0 \\
24.05.2014  &C  &2,2,2,2&       18.07.2015 &D  &33,0,33,32      & 05.11.2015 &D  &30,0,30,30    & 26.08.2016 &D  &25,26,59,0 \\
15.06.2014  &B  &5,5,5,5&       19.07.2015 &D  &26,0,27,27      & 06.11.2015 &D  &31,0,33,32    & 27.08.2016 &D  &9,9,9,9 \\
24.06.2014  &C  &2,2,2,2&       20.07.2015 &D  &22,0,24,24      & 07.11.2015 &D  &28,0,32,32    & 28.08.2016 &D  &14,14,12,13 \\
26.06.2014  &C  &2,2,2,2&       21.07.2015 &D  &20,0,20,20      & 07.11.2015 &E  &1,1,1,1       & 29.08.2016 &D  &8,9,9,9  \\
06.07.2014  &B  &5,5,5,5&       22.07.2015 &D  &23,0,24,24      & 11.11.2015 &D  &46,0,46,45    & 30.08.2016 &D  &45,54,53,53 \\
21.07.2014  &B  &5,5,5,5&       23.07.2015 &D  &18,0,22,20      & 12.11.2015 &D  &43,0,42,41    & 19.09.2016 &B  &3,3,3,3  \\
22.07.2014  &B  &5,5,5,5&       23.07.2015 &B  &5,5,5,5         & 12.11.2015 &E  &1,1,1,1       & 20.09.2016 &B  & 5,5,5,5  \\
25.07.2014  &B  &5,5,5,5&       11.08.2015 &E  &1,1,1,1         & 13.11.2015 &D  &3,0,2,2       & 21.09.2016 &B  &5,5,5,5  \\
28.07.2014  &B  &5,5,5,5&       18.08.2015 &D  &28,0,28,28      & 03.12.2015 &D  &15,0,16,18    & 22.09.2016 &D  &0,0,9,9  \\
29.07.2014  &B  &10,10,10,10&   24.08.2015 &A  &6,0,6,6         & 04.12.2015 &E  &1,1,1,1       & 23.09.2016 &D  &44,44,42,23 \\
28.09.2014  &D  &13,17,19,23&   24.08.2015 &D  &3,2,2,2         & 05.12.2015 &D  &2,3,3,3       & 24.09.2016 &D  &31,30,30,22 \\
29.09.2014  &D  &12,16,17,19&   26.08.2015 &D  &17,0,20,17      & 05.12.2015 &E  &1,1,1,1       & 25.09.2016 &B  & 5,5,5,5  \\
30.09.2014  &D  &19,20,22,23&   27.08.2015 &D  &19,0,21,20      & 13.12.2015 &D  &2,0,3,2       & 26.09.2016 &D  &45,45,43,42 \\
30.09.2014  &E  &1,1,1,1&       28.08.2015 &D  &22,0,23,21      & 14.12.2015 &D  &29,0,30,30    & 27.09.2016 &D  &42,42,41,37 \\
20.02.2015  &A  &2,2,2,2&       08.09.2015 &D  &45,0,45,45      & 14.12.2015 &E  &1,1,1,1       & 28.09.2016 &D  &50,46,47,49 \\
20.05.2015  &A  &2,2,2,2&       12.09.2015 &D  &2,0,2,2         & 30.06.2016 &D  &0,20,20,20    & 29.09.2016 &D  &49,49,50,42 \\
14.06.2015  &C  &2,2,2,2&       14.09.2015 &D  &23,0,27,29      & 01.07.2016 &D  &0,14,14,13    & 30.09.2016 &D  &38,35,34,36 \\
29.06.2015  &B  &10,10,10,10&   15.09.2015 &D  &28,0,28,30      & 08.08.2016 &D  &15,15,30,0    & 01.10.2016 &D  &2,2,2,2 \\
30.06.2015  &B  &10,10,10,10&   03.11.2015 &D  &21,0,21,22      & 09.08.2016 &D  &25,25,51,0    &            &   &   \\\hline
\end{tabular} \\
\footnotesize
A: 50/70-cm Schmidt Telescope at  National Astronomical Observatory Observatory, Rozhen, Bulgaria   \\
B: 1.3-m Skinakas Observatory, Crete, Greece \\
C: 2m RCC, National Astronomical Observatory, Rozhen, Bulgaria   \\
D: 60-cm Cassegrain Telescope at Astronomical Observatory Belogradchik, Bulgaria \\
E: 60-cm Cassegrain Telescope, Astronomical Station Vidojevica (ASV), Serbia \\
F: 70-cm meniscus telescope at Abastumani Observatory, Georgia  \\
G: 35.6 cm Telescope at Observatorio Astronomico Las Casqueras, Spain   \\
H: 2.3-m Bok Telescope and 1.54-m Kuiper Telescope at Steward Observatory, Arizona, USA   \\

\end{table*}


\end{document}